\documentclass[a4paper,12pt,english]{article}

\pdfoutput=1

\usepackage[bindingoffset=0.3cm,textheight=22.5cm,hdivide={2cm,*,2cm}, vdivide={*,22cm,*}]{geometry}

\usepackage{amsmath,amsfonts,amssymb,babel,slashed,color,empheq,tensor}
\usepackage[svgnames]{xcolor}
\usepackage[pdftex,breaklinks,colorlinks=true,urlcolor=RoyalBlue,%
linkcolor=blue,citecolor=blue]{hyperref}

\usepackage[utf8,latin1]{inputenc}
\usepackage{graphicx}
\usepackage{dcolumn}

\usepackage{mathabx}
\usepackage{mathrsfs}  
\usepackage{cases}
\usepackage{bm}
\newcommand{\qm}[1]{``#1''}
\usepackage{hyperref}
\usepackage{stackengine,scalerel}
\hypersetup{colorlinks, linkcolor={red},citecolor={blue},urlcolor={blue}}  
\usepackage{bbold}
\usepackage{authblk}

\usepackage[numbers,square,comma,sort&compress]{natbib}
\usepackage{adjustbox}

\makeatletter

\def\R{{\mathbb R}} \def\C{{\mathbb C}} 
 \def\one{\mbox{1 \kern-.59em {\rm l}}}

\newcommand{\Tr}{\mathrm{Tr}}

\newcommand{\End}{\mathrm{End}}

  \def\cC{{\cal C}} 
   
 \def\cH{{\cal H}}  
   
\def\cM{{\cal M}}   
 \def\cQ{{\cal Q}}

 \def\a{\alpha}  \def\b{\beta}
 \def\g{\gamma} 
 \def\d{\delta} 

\newcommand{\eq}[1]{(\ref{#1})}
\newcommand{\del}{\partial}
\def\nn{\nonumber}

 \allowdisplaybreaks[3]

\makeatother

\renewcommand{\title}[1]{\vspace{10mm}\noindent{\Large{\bf

#1}}\vspace{8mm}} \newcommand{\authors}[1]{\noindent{\large

#1}\vspace{5mm}} \newcommand{\address}[1]{{\itshape #1\vspace{2mm}}}

\begin{document}

 \begin{flushright}
  UWThPh-2022-10 
 \end{flushright}
\begin{center}
\title{ {\Large On the propagation across the big bounce in an open \\[1ex] quantum FLRW  cosmology} }

\vskip 3mm

\authors{Emmanuele Battista\footnote{emmanuele.battista@univie.ac.at, emmanuelebattista@gmail.com}, Harold C. Steinacker\footnote{harold.steinacker@univie.ac.at}}

 \vskip 2mm

  \address{ 
{\it Faculty of Physics, University of Vienna\\
 Boltzmanngasse 5, A-1090 Vienna, Austria  }  
   }

\bigskip


\textbf{Abstract}
\vskip 3mm

\begin{minipage}{14.3cm}%

The propagation of a  scalar field in an open FLRW bounce-type quantum spacetime is examined, 
which arises within the framework of the IKKT matrix theory.
In the first part of the paper, we employ general-relativity tools to study  null and timelike geodesics at the classical level. This analysis reveals that  massless and massive non-interacting particles can  
travel across the big bounce. We then exploit quantum-field-theory techniques to  evaluate the scalar field propagator. In the late-time regime, we find that it resembles the standard Feynman propagator of flat Minkowski space, whereas for early times it governs the propagation across the big bounce and gives rise to a well-defined correlation between two points on opposite sheets of the spacetime.

\end{minipage}

\end{center}

{
  \hypersetup{linkcolor=black}
  \tableofcontents
}

\section{Introduction}

Bouncing cosmology  has been proposed in the literature either as alternatives or completions of the inflationary paradigm. In Refs. \cite{Klinkhamer:2019dzc,Klinkhamer2019b,Wang2021a,Battista2020a}, it has been shown that the big-bang singularity can be regularized by employing a   slight modification of Einstein theory allowing for degenerate metrics. The 3-dimensional submanifold of the spacetime  where the metric is degenerate represents a spacetime defect and it has been recently suggested that its origin can be explained within the IIB matrix model \cite{Steinacker:2017vqw,Steinacker:2017bhb,Sperling:2019xar,Klinkhamer2020a}.  Universes having
a bouncing-like behavior have been addressed  in the context of $f(R)$ theories  in   Refs. \cite{Odintsov2019,Odintsov2020a,Odintsov2020b} and in Gauss-Bonnet modified gravity in Ref. \cite{Escofet2015}. Furthermore, the authors of Refs. \cite{Easson2011,Ijjas2016,Ijjas2016b} have constructed classical nonsingular bouncing models by means of  generalized cubic Galileon theories, which permit to realize a pattern where the null energy condition  is violated without introducing  ghost and gradient instabilities. Another viable theory is represented by the matter bounce scenario \cite{Cai2012}. Last, the  bounce mechanism has been  investigated  in the framework of loop quantum cosmology  \cite{Corichi2007, Bojowald2008, Bojowald2009, Wilson-Ewing2012} and string cosmology \cite{Veneziano2003,Gasperini2007}. For a review on bouncing cosmologies we refer the reader to Refs. \cite{Novello2008,Cai2014,Battefeld2014,Banerjee2022}.  

Recently, solutions of matrix models have been found \cite{Sperling:2019xar,Steinacker:2017vqw,Steinacker:2017bhb}, which can be interpreted as 3+1-dimensional quantum geometries\footnote{See also e.g. \cite{Brahma:2021tkh,Hatakeyama:2019jyw,Stern:2014aqa,Chaney:2015ktw,Kim:2011ts,Klinkhamer:2020wct} for related work.} describing an effective FLRW cosmology with a big bounce (BB). The underlying model is known as IKKT model \cite{Ishibashi:1996xs}, which has been proposed as a constructive definition of (some corner of) string theory. These solutions or backgrounds have an intrinsic quantum structure, with spacetime uncertainty or "fuzzyness" akin to  quantum mechanical phase space. In this framework, a classical spacetime geometry is recovered in the semi-classical or IR regime, while the quantum structure of geometry becomes important only in the UV regime, i.e. at very short distances. In particular, the singularity of classical geometry at the BB is completely under control. Fluctuations of such a  background lead to fields, including scalar fields, gauge fields, and gravitons; in fact for the geometry given in \cite{Sperling:2019xar},  a whole tower of higher-spin gauge fields arises, which is described by a ghost-free 
higher-spin gauge theory \cite{Steinacker:2019awe}.  

In particular, 
the framework of matrix models allows to study the physics on such backgrounds with a BB. This study was initiated in \cite{Karczmarek:2022ejn} on a 1+1-dimensional toy model, which allowed to compute the propagator for the global geometry including the BB. Furthermore, the Bogoljubov coefficients which govern the asymptotic properties of fields propagating in and out of the BB were obtained.

In the present paper, we extend the results of \cite{Karczmarek:2022ejn} to the case of 3+1 dimensions.
We obtain explicitly the fluctuation modes of a scalar field on the 3+1-dimensional FLRW background in the semi-classical regime, paying special attention to the asymptotic regimes of late times and close to the BB.
This then allows to compute the propagator explicitly, using a path integral approach provided by the underlying matrix model framework. Remarkably, the propagator is found to be regular at the BB, and 
has  the local structure of a Feynman propagator at late times. This means that physical modes can propagate across the BB singularity in a well-defined way, and some of their structure will survive.

 The main message of this paper is that  matrix models provide a suitable framework for quantum geometry, which allows to address and resolve  the singularities which arise in the framework of general relativity. Moreover, the present example demonstrates how a time evolution and a 3+1-dimensional causal structure can emerge from the underlying matrix model, which has no a priori notions of space and time.
 It should be emphasized that the dynamics is governed here by the matrix model, which is different from general relativity, at least at the classical level. As demonstrated in \cite{Steinacker:2021yxt}, the Einstein-Hilbert action does arise upon including 1-loop effects in the IKKT model, under suitable assumptions. This will of course affect some of the results of the present paper, however we expect that the qualitative features of the present classical analysis will also apply after including quantum effects.

The paper is morally divided into two parts, a classical and a quantum one. The two parts are consistent with each other. In the classical part, we study some aspects of the present FLRW geometry, with special emphasis on the near-BB regime. In particular, we elaborate the  geodesics, and show that they extend smoothly across the BB. The BB singularity is found to be rather \qm{mild} in a sense that will be explained below. The propagator is obtained in the quantum setting, by computing explicitly the (free) path integral of modes as defined by the matrix model. At late times, we recover again the standard Feynman propagator with the appropriate $i\varepsilon$ structure. At early times near the BB, the propagator also turns out to be well-defined, and allows to study the propagation of  scalar particles across the BB. This result agrees with the classical analysis regarding null and timelike geodesic.

\emph{Notations.} We use metric signature  $(-,+,+,+)$. Greek indices take values  $0,1,2,3$. The flat metric is indicated by $\eta^{\alpha \beta }=\eta_{\alpha \beta }={\rm diag}(-1,1,1,1)$.

 \section{The background geometry}

We recall \cite{Sperling:2019xar} that  the background  $\cM^{3,1}$ under consideration  can be described semi-classically as a projection of fuzzy $H^4_n$, which is obtained from five matrices $X^a \sim x^a$ interpreted as quantizations of five embedding functions 
\begin{align}
   x^a: \quad H^4 \hookrightarrow \R^{4,1}
\end{align}
where $a=0,...,4$.
A convenient parametrization of 
this 4-dimensional hyperboloid is as follows 
\begin{align}
 \begin{bmatrix}
  x^0 \\ x^1 \\ x^2 \\x^3 \\ x^4
 \end{bmatrix}
= R \begin{bmatrix}
 \cosh(\eta) 
\begin{pmatrix}
\cosh(\chi) \\
\sinh(\chi)\sin(\theta) \cos(\varphi) \\
\sinh(\chi)\sin(\theta) \sin(\varphi) \\
\sinh(\chi)\cos(\theta)
\end{pmatrix} \\
\sinh(\eta) 
\end{bmatrix}, \ 
\label{embedding-4d-hyperboloid}
\end{align}   
for $\eta\in\R$. Note that $\chi$ can be restricted to be positive.
Projecting this along the $x^4$ axis leads to a 2-sheeted cover of
the following region
\begin{align}
    x_\mu x^\mu \leq -R^2,
\end{align}
where the upper sheet (\qm{post-BB}, corresponding to $x^4>0$) is
covered by $\eta > 0$, while the lower sheet (\qm{pre-BB}, corresponding to $x^4<0$) is covered by $\eta < 0$.
The BB separates these sheets, and corresponds to $x_\mu x^\mu = -R^2$.
This leads to the following parametrization of $\cM^{3,1}$
\begin{align}
 \begin{pmatrix}
  x^0 \\ x^1 \\ x^2 \\x^3 
 \end{pmatrix}
= R \cosh(\eta) 
\begin{pmatrix}
\cosh(\chi) \\
\sinh(\chi)\sin(\theta) \cos(\varphi) \\
\sinh(\chi)\sin(\theta) \sin(\varphi) \\
\sinh(\chi)\cos(\theta)
\end{pmatrix} \ .
\label{embedding-3d-hyperboloid}
\end{align} 
Note that the flow of time will be along increasing $\eta$ on both sheets; this arises from the $i\varepsilon$  regularization discussed in Sec. \ref{sec:path-integral}.

In principle, we can restrict $\chi$ to be either positive or negative on either sheet. However, we will see in Secs. \ref{Sec:scalar-modes} and \ref{sec:path-integral} that it is convenient to choose $\chi>0$.

\subsection{Effective metric}
\label{sec:metric}

To understand the effective metric on $\cM^{3,1}$,
we recall \cite{Sperling:2019xar} that the background solution $T^\mu$ of the matrix model leads to the following kinetic term 
\begin{align}
S[\phi] =  - \Tr [T^\mu,\phi][T_\mu,\phi] 
\label{kinetic-term}
\end{align}
which governs {\em all} fluctuations in the matrix model. Using the semi-classical relation $[T^\mu,.] \sim i\{t^\mu,.\}$ in terms of Poisson brackets (here $t^\mu$ represents the semiclassical limit of $T^\mu$) and recalling $\{t^\mu,x^\nu\} = \sinh(\eta)\d^\mu_\nu$, this can be rewritten uniquely in the standard form
 \cite{Sperling:2019xar,Steinacker:2010rh}
\begin{align}
S[\phi] =  - \Tr [T^\mu,\phi][T_\mu,\phi] 
\sim \int d^4 x\,\sqrt{|G|}G^{\mu\nu}\del_\mu\phi \del_\nu \phi \ 
\label{scalar-action-metric}
\end{align}
where
\cite{Sperling:2019xar}
\begin{align}
   G^{\mu\nu} &= \vert  \sinh(\eta) \vert^{-3}\, \g^{\mu\nu},
   \qquad  \g^{\a\b} = 
   \sinh^2(\eta) \eta^{\a\b},
   \label{eff-metric-G}
\end{align}
dropping some irrelevant constant (here $\g^{\mu\nu}$ is an auxiliary metric which is relevant for the torsion).
This metric can be recognized as  $SO(3,1)$-invariant FLRW metric, 
\begin{align}
 d s^2_G = G_{\mu\nu} d x^\mu d x^\nu 
   &= -R^2 \vert \sinh(\eta) \vert^3 d \eta^2 + R^2 \vert \sinh(\eta)\vert \cosh^2(\eta)\, d \Sigma^2 \ \nn\\
   &= -d t^2 + a^2(t)d\Sigma^2 \, .
   \label{eff-metric-FRW}
\end{align}
where 
\begin{align}
    d\Sigma^2 = d\chi^2 + \sinh^2\chi (d\theta^2 + \sin^2 \theta d\varphi^2),
    \label{dSigma2}
\end{align}
is the invariant length element on the space-like hyperboloids $H^3$ (with $-\infty \leq \chi <\infty$, $0 \leq \theta < \pi$, $0 \leq \varphi <2\pi$). Equivalently, we can write
\begin{align}
  d\Sigma^2=\dfrac{dr^2}{1+r^2} + r^2  (d\theta^2 + \sin^2 \theta d\varphi^2),
\end{align}
with
\begin{equation}
    r = \sinh \chi.
    \label{r-sinh-chi}
\end{equation}

From Eq. \eqref{eff-metric-FRW}, we can read off the cosmic scale parameter $a(\eta)$ and the relation linking the differentials $dt$ and $d\eta$, i.e.,
\begin{align}
\vert a(\eta) \vert &=  R \cosh(\eta) \vert \sinh(\eta)\vert^{1/2}  ,  
\label{a-eta}
\\
d t &=  R \vert \sinh(\eta)\vert^{3/2} d\eta.  
\label{dt-squared}
\end{align}
For late times, we have
\begin{equation}
  a^2(t)  \ \stackrel {t\to\infty}{\sim}  \  R^2\vert \sinh(\eta)\vert^3 \ .
\end{equation}
This is a reasonable FLRW cosmology given the simplicity of the model, which is
asymptotically coasting at late time
with  $a(t) \sim \frac 32 t$, cf. Refs. \cite{Kolb:1989bg,Melia:2011fj}. Note that it arises directly from the matrix model, without using or assuming general relativity. Last, it is worth mentioning that the scale factor \eqref{a-eta} can be either odd or even (the odd solution, in particular, may
be of interest as discussed in Ref. \cite{Boyle2018}).

\section{Classical analysis of the FLRW spacetime}\label{Sec:classical-analysis}

In this section, we perform a classical investigation of the FLRW geometry \eqref{eff-metric-FRW}. Curvature invariants are considered in Sec. \ref{Sec:Curvature-invariants}, whereas null and timelike geodesics are studied in Sec. \ref{Sec:geodesics}. We conclude the section with the analysis of some cosmological observables (see Sec. \ref{Sec:Cosmol-obs}).

\subsection{Curvature invariants} \label{Sec:Curvature-invariants}

In order to describe the behaviour of the spacetime near the BB, we begin our analysis with the investigation of some curvature invariants. Starting from the metric \eqref{eff-metric-FRW}, we find that the Kretschmann scalar is
\begin{equation}\label{kretschmann-invariant}
    R_{\mu \nu \rho \sigma}R^{\mu \nu \rho \sigma}= \dfrac{3}{32R^4 \sinh^{10}\left(\eta\right) } \left[171-60 \cosh\left(2 \eta\right)+25\cosh\left(4 \eta\right)\right],
\end{equation}
 the squared Ricci tensor reads as
\begin{align}
    R_{\mu \nu}R^{\mu \nu}&=\dfrac{3}{512R^4 \sinh^{10}\left(\eta\right)\cosh^{4}\left(\eta\right) }\Bigl[1635-488 \cosh\left(6\eta\right)+97 \cosh\left(8\eta\right) -384 \sinh^4\left(\eta\right)
    \nonumber \\
    & +4 \cosh\left(4\eta\right) \left(287-288\sinh^4\left(\eta\right)\right) +8\cosh\left(2\eta\right) \left(320\sinh^4\left(\eta\right)-91\right)\Bigr], 
    \label{squared-Ricci-tensor}
\end{align}
whereas the (topological) Euler invariant is \cite{Obukhov1995,Cherubini2002,Steane2021}
\begin{align}
\label{euler-scalar}
    {}^{\star}R^{\star}_{\mu \nu \rho \sigma}R^{\mu \nu \rho \sigma}=  \dfrac{3}{4R^4 \sinh^{10}\left(\eta\right)\cosh^{2}\left(\eta\right)} \left[11-12\cosh\left(2\eta\right)+9\cosh\left(4\eta\right)-32 \sinh^4\left(\eta\right)\right],
\end{align}
the star indicating  the duality operation. It is clear that the scalars \eqref{kretschmann-invariant}--\eqref{euler-scalar} blow up at the BB, i.e., at $\eta=0$.

\subsection{Null and timelike geodesics}\label{Sec:geodesics}

In this section, we investigate  null and timelike geodesics of the FLRW geometry having  $\theta$ and $\varphi$ constant. 

Null geodesics can be described in terms of the function $\chi(\eta)$. We seek  a solution which reaches the BB at $\eta=0$ after having travelled toward it for $\eta<0$ and away from it when $\eta>0$. Therefore,   it follows from Eq. \eqref{eff-metric-FRW} that the motion  in the outward $\chi$-direction is parametrized by  the differential equation
\begin{equation}
    \dfrac{d \chi}{d \eta} = \vert \tanh \eta \vert,
\end{equation}
which, with the boundary condition $\chi(\eta=0)=0$, leads to 
\begin{equation}
    \chi(\eta) = \left \{ 
\begin{array}{rl}
& \;\;\;   \log \left(\cosh \eta\right),  \quad \, \eta \geq 0,\\
& -\log \left(\cosh \eta\right), \, \quad \eta<0.
\end{array}
\right.
\label{chi-of-eta-solution}
\end{equation}

\begin{figure*}[t!]
    \centering
    \includegraphics[scale=0.72]{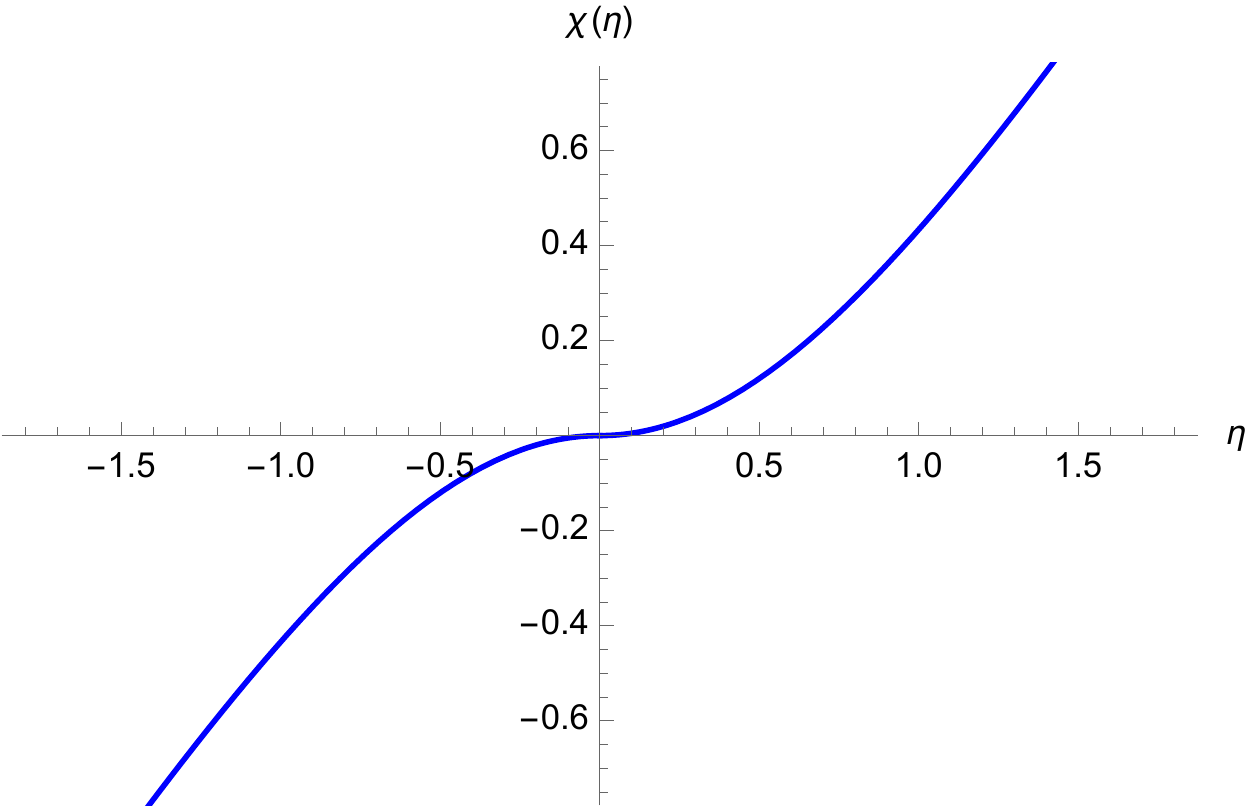}
    \caption{Null geodesic motion having $\theta$ and $\varphi$ constants (cf. Eq. \eqref{chi-of-eta-solution}). It is clear that the function $\chi(\eta)$ is continuous at the BB.}
    \label{fig:null-geodesic}
\end{figure*} 

\begin{figure*}[t!]
    \centering
    \includegraphics[scale=0.72]{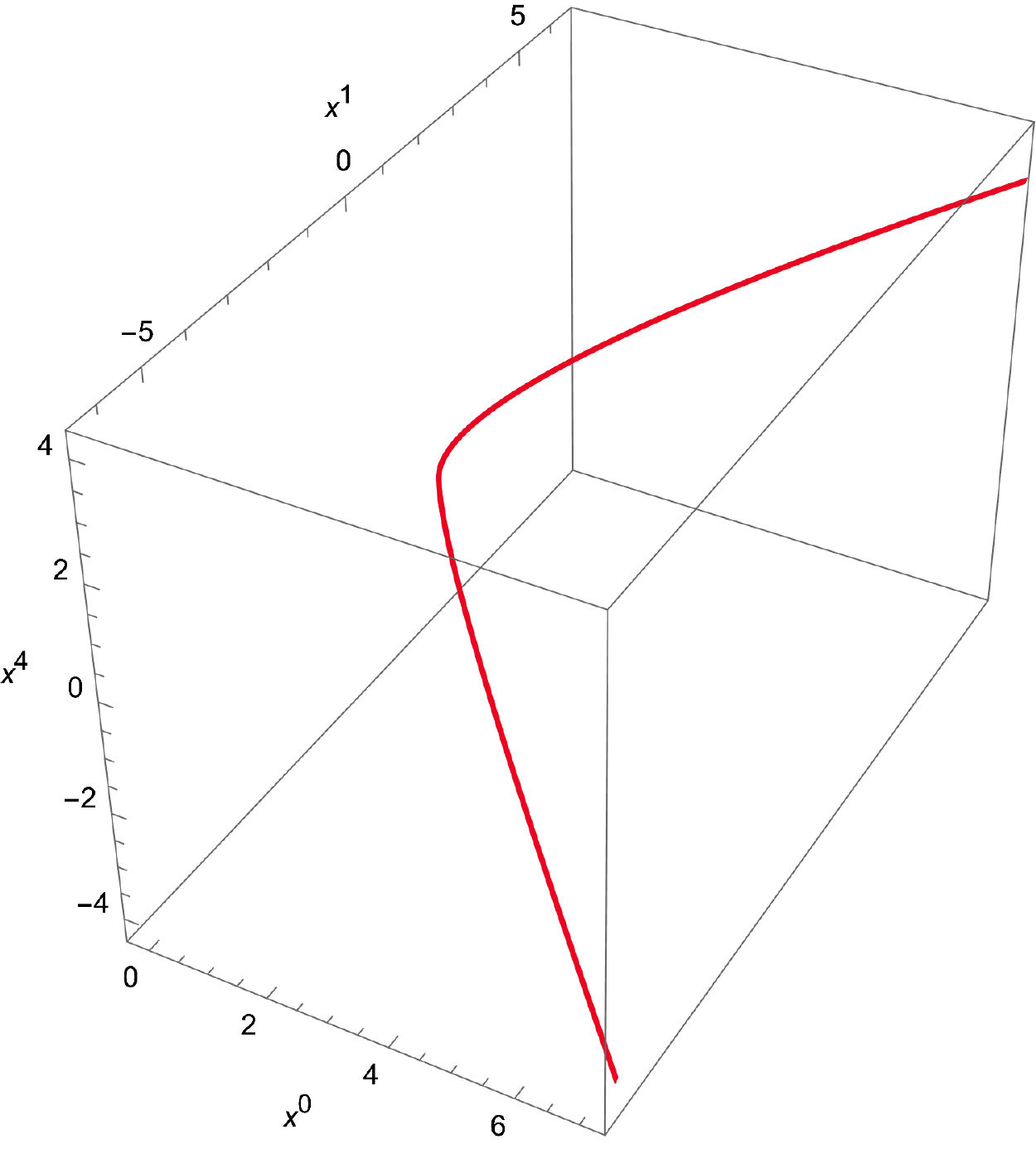}
    \caption{Null geodesic motion obtained by means of Eqs.  \eqref{embedding-4d-hyperboloid} and \eqref{chi-of-eta-solution}. The following values have been chosen: $R=1$, $\theta=\pi/2$, and $\varphi=0$. }
    \label{fig:null-geodesic-3D-plot}
\end{figure*}

The behaviour of the solution \eqref{chi-of-eta-solution} is shown in   Fig. \ref{fig:null-geodesic}, whereas Fig. \ref{fig:null-geodesic-3D-plot} represents the plot obtained by means of the embedding functions \eqref{embedding-4d-hyperboloid}. 
Having obtained a    continuous  geodesic solution which can be extended uniquely   at $\eta=0$,   we can conclude that light (and hence the physical information) can 
travel across the BB, despite the singularity occurring in the  invariants   \eqref{kretschmann-invariant}--\eqref{euler-scalar}.

It is interesting to work out the solution $\chi = \chi (t)$  of null geodesics (with $\theta$ and $\varphi$ constant) for early times, i.e., when $t \to t_0$ (with $t_0 \in \mathbb{R}$) or, equivalently $\eta \to 0$. First of all, from Eq. \eqref{dt-squared} jointly with the condition $t(\eta=0)=t_0$, we obtain
\begin{align}
    t(\eta) -t_0 & \; \overset{\eta \to 0}{\sim} \; \dfrac{2}{5} R \eta \vert \eta \vert^{3/2} =\left \{ 
\setlength{\tabcolsep}{10pt} 
\renewcommand{\arraystretch}{1.5}
\begin{array}{rl}
&    \dfrac{2}{5} R \eta^{5/2},   \qquad \qquad \,\,\, \eta \geq 0,\\
& \dfrac{2}{5} R \eta \left(-\eta\right)^{3/2},  \qquad \eta<0.
\end{array}
\right.
\end{align}
The inversion of the above function yields 
 \begin{align}
     \eta(t) & \; \overset{t \to t_0}{\sim} \left \{ 
\setlength{\tabcolsep}{10pt} 
\renewcommand{\arraystretch}{1.5}
\begin{array}{rl}
& \;\;\,\,   \left(\dfrac{5}{2R}\right)^{2/5}  \left(t-t_0\right)^{2/5},   \qquad  t \geq t_0,\\
&  -\left(\dfrac{5}{2R}\right)^{2/5}  \left(t-t_0\right)^{2/5},  \qquad t<t_0,
\end{array}
\right.
\label{eta-of-t-early-times}
 \end{align}
where we note that $\eta (t=t_0)=0$. 

\begin{figure*}[t!]
    \centering
    \includegraphics[scale=0.72]{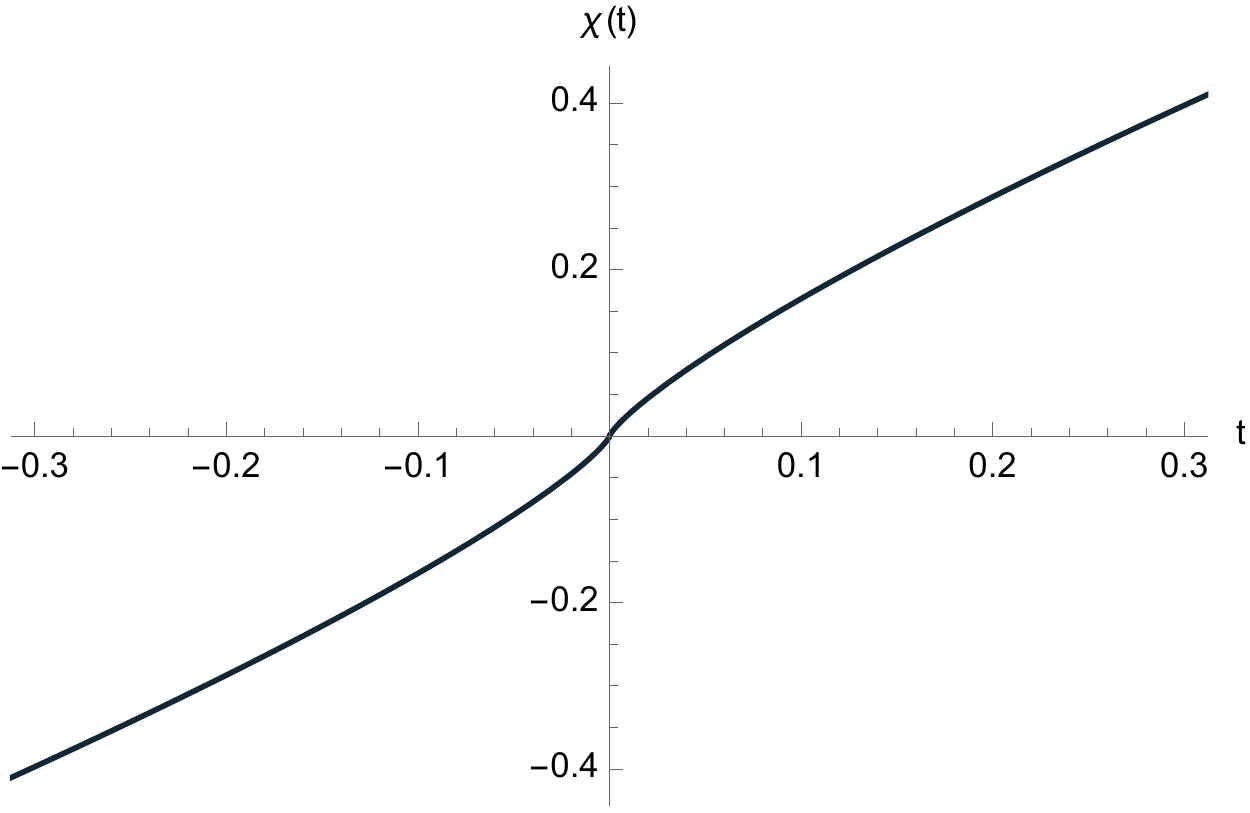}
    \caption{The function \eqref{chi-of-t-null-geodesics-small-times} describing early-time null geodesic having $\theta$ and $\varphi$ constant. We have chosen $R=1$ and $t_0=0$.}
    \label{fig:null-geodesic-small-times}
\end{figure*}

We are now ready to obtain the expression of the cosmic scale factor valid near the BB. Indeed, by means of Eqs. \eqref{a-eta} and \eqref{eta-of-t-early-times}, we have 
 \begin{align}
    a(t) & \; \overset{t \to t_0}{\sim} \left \{ 
\setlength{\tabcolsep}{10pt} 
\renewcommand{\arraystretch}{1.5}
\begin{array}{rl}
& \;\;\,   R\left(\dfrac{5}{2R}\right)^{1/5}  \vert t-t_0 \vert^{1/5},   \qquad  t \geq t_0,\\
&  -R\left(\dfrac{5}{2R}\right)^{1/5}  \vert t-t_0 \vert^{1/5},  \qquad t<t_0,
\end{array}
\right.
&=R\left(\dfrac{5}{2R}\right)^{1/5}  ( t-t_0 )^{1/5},
\label{scale-factor-early-times}
 \end{align}
which vanishes at $t=t_0$. We note that $a(t)$ is positive (resp. negative) for $t>t_0$ (resp. $t<t_0$).  The sign of $a(t)$ drops out in the metric \eqref{eff-metric-FRW}, but the above choice has no cusp. The equation of null geodesics (with $\theta$ and $\varphi$ constant) can be written in the equivalent form
\begin{align}\label{null-geodesic-in-terms-of-t}
     \dfrac{d \chi}{d t} = \dfrac{1}{\vert a(t)\vert},
\end{align}
which in view of \eqref{scale-factor-early-times} gives the desired early-time solution
 \begin{align}
        \chi(t) & \; \overset{t \to t_0}{\sim} \left \{ 
\setlength{\tabcolsep}{10pt} 
\renewcommand{\arraystretch}{1.5}
\begin{array}{rl}
& \;\;\,   \dfrac{1}{2}\left(\dfrac{5}{2R}\right)^{4/5}  \left( t-t_0 \right)^{4/5},   \qquad  t \geq t_0,\\
&  -\dfrac{1}{2}\left(\dfrac{5}{2R}\right)^{4/5}  \left( t-t_0 \right)^{4/5},  \qquad t<t_0,
\end{array}
\right.  
     \label{chi-of-t-null-geodesics-small-times}
\end{align}
see Fig. \ref{fig:null-geodesic-small-times}. The above solution can also be obtained if we first expand Eq. \eqref{chi-of-eta-solution} about $\eta = 0$, and then exploit Eq. \eqref{eta-of-t-early-times}.

At this stage,  we analyze the  timelike geodesics of a non-comoving observer. Hence, let
\begin{align}
    u^\alpha = \dfrac{dx^\alpha(\tau)}{d \tau},
\end{align}
denote the unit timelike four-velocity vector of such observer, whose proper time is indicated with $\tau$. If we suppose, like before, that the observer moves along the direction of constant $\theta$ and $\varphi$, then from the normalization condition
\begin{align}
    g_{\alpha \beta} u^\alpha u^\beta = -1,
\end{align}
we obtain 
\begin{align}
    R^2 \vert \sinh \eta \vert^3 \left(\dfrac{d \eta(\tau)}{d\tau}\right)^2= 1+a^2(\eta)\left(\dfrac{d \chi(\tau)}{d\tau}\right)^2.
    \label{timelike-geod-calculation}
\end{align}
Furthermore, if we consider the $\chi$-translational Killing vector $\xi^\alpha$, the conserved momentum $\Pi$ of the non-comoving observer is given by
\begin{equation}
    \Pi= \xi^\alpha u_\alpha = a^2(\eta)\dfrac{d \chi(\tau)}{d\tau}.
\end{equation}
\begin{figure*}[t!]
    \centering
    \includegraphics[scale=0.72]{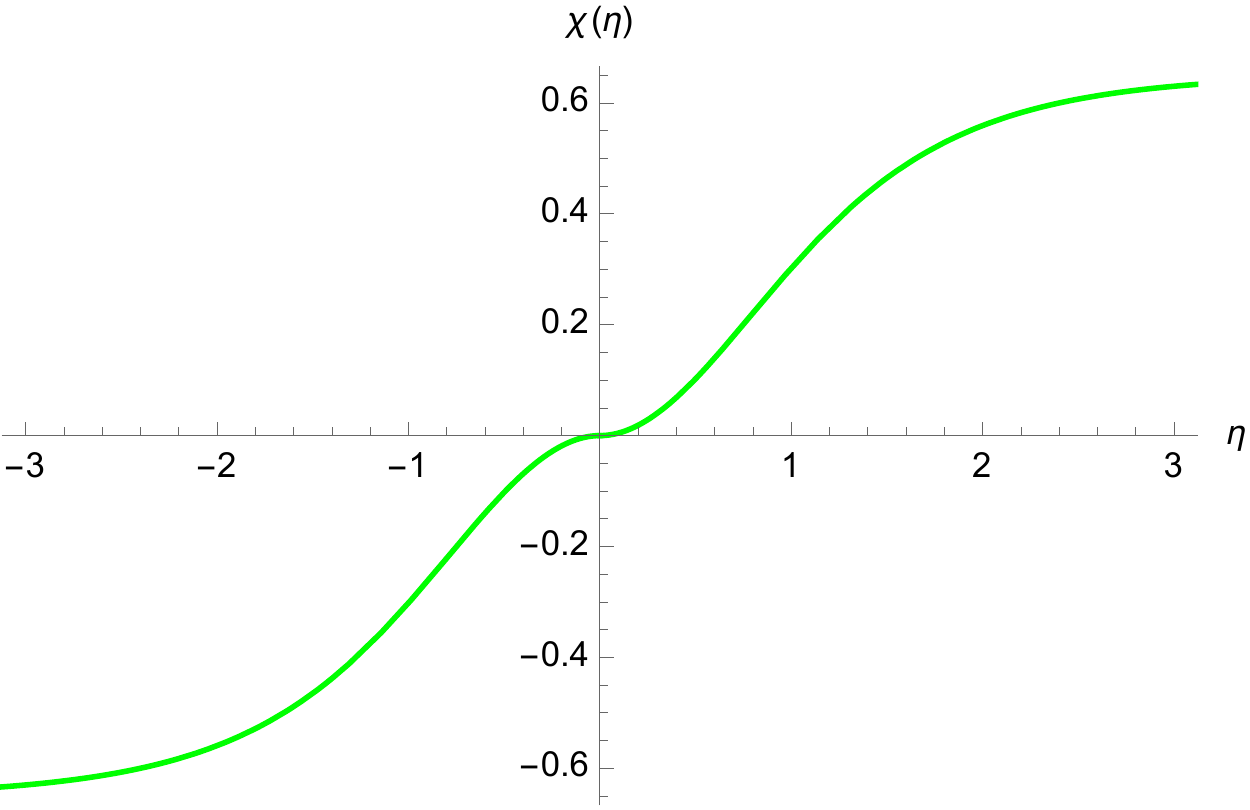}
    \caption{Timelike geodesic of the non-comoving observer obtained by solving numerically Eq. \eqref{timelike-geod-equation} with the  boundary condition $\chi(\eta=0)=0$. The following constants have been  chosen: $R=1$ and $\Pi=1$.}
    \label{fig:timelike-geodesic}
\end{figure*}
The above equation gives a relation to express $d \chi(\tau)/d\tau$ in terms of $\Pi$ and $a(\eta)$, which  can be  exploited in Eq. \eqref{timelike-geod-calculation}. In this way, after some manipulations, we end up with the geodesic equation in terms of the function $\chi(\eta)$, i.e.,  
\begin{align}\label{timelike-geod-equation}
    \dfrac{d \chi(\eta)}{d\eta} &= \dfrac{d\chi(\tau)/d\tau}{d\eta(\tau)/d\tau}=\dfrac{\vert \tanh \eta \vert }{\sqrt{1+a^2(\eta)/\Pi^2}}.
\end{align}

We have solved numerically Eq. \eqref{timelike-geod-equation} with the  boundary condition $\chi(\eta=0)=0$. As it is clear from Fig. \ref{fig:timelike-geodesic}, the non-comoving observer can 
travel across
the BB, likewise a massless particle.

\begin{figure*}[t!]
    \centering
    \includegraphics[scale=0.72]{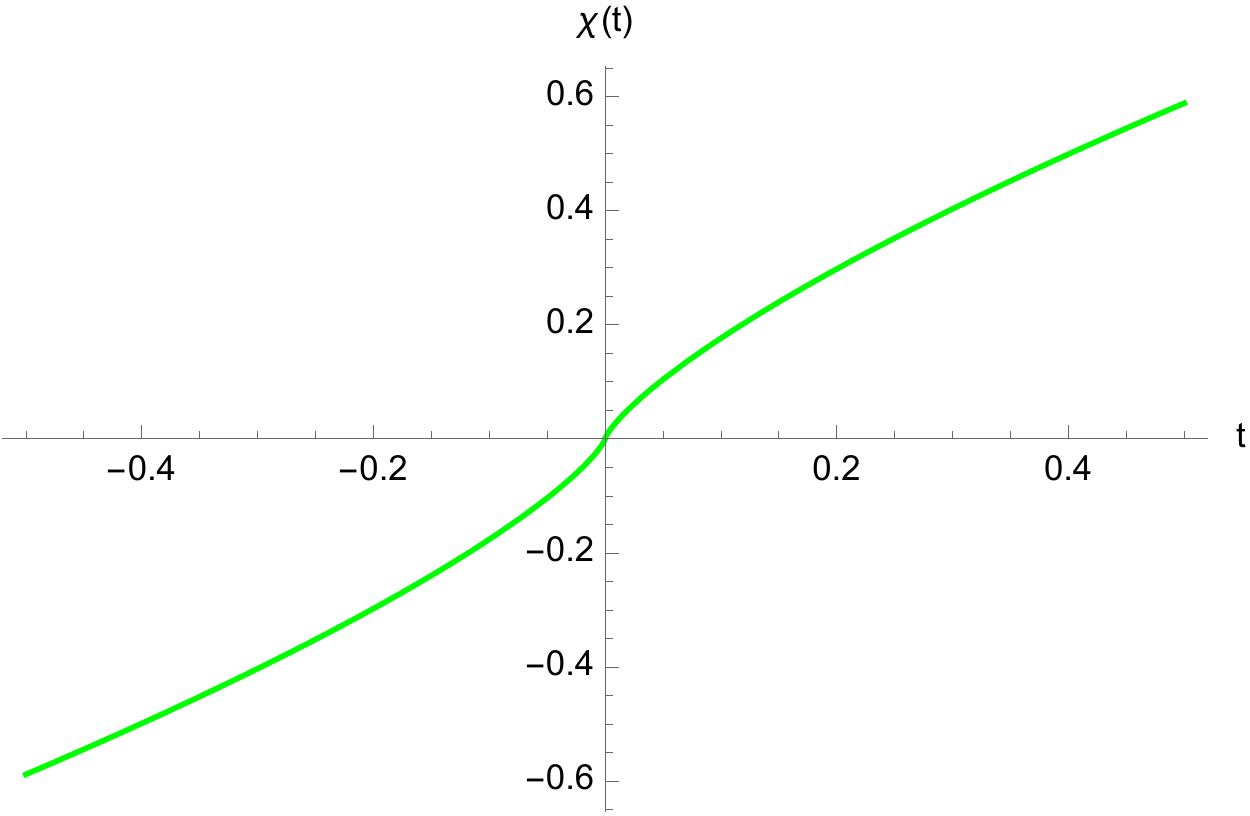}
    \caption{The function \eqref{chi-of-t-timelike-geodesics-small-times} describing the early-time timelike geodesic of the non-comoving observer with $t_0=0$, $R=1$, and $\Pi=1$ and the  boundary condition $\chi(t=0)=0$.}
    \label{fig:timelike-geodesic-small-time}
\end{figure*}
We can provide the analytic solution for the timelike geodesic of the non-comoving observer moving in the $\chi$-direction in the regime of small times. First of all,  it is easy to see that the equation of the timelike geodesic of the non-comoving observer in terms of the cosmic time variable is
\begin{align}\label{timelike-geodesic-in-terms-of-t}
    \dfrac{d\chi(t)}{dt}=\dfrac{1}{\vert a(t)\vert}\dfrac{1}{\sqrt{1+a^2(t)/\Pi^2}}.
\end{align}
Bearing in mind Eq. \eqref{scale-factor-early-times} and considering, for simplicity, $t_0=0$, $R=1$, and $\Pi=1$, the solution of Eq. \eqref{timelike-geodesic-in-terms-of-t} with the boundary condition $\chi(t=0)=0$ reads as
 \begin{align}
        \chi(t) & \; \overset{t \to 0}{\sim} \left \{ 
\setlength{\tabcolsep}{10pt} 
\renewcommand{\arraystretch}{2.0}
\begin{array}{rl}
& \;\;\,   \dfrac{5}{3}\left[2-2\sqrt{1+t^{2/5}}+t^{2/5}\sqrt{1+t^{2/5}}\right],   \qquad  t \geq 0,\\
&  -\dfrac{5}{3}\left[2-2\sqrt{1+t^{2/5}}+t^{2/5}\sqrt{1+t^{2/5}}\right],  \qquad t<0.
\end{array}
\right.  
     \label{chi-of-t-timelike-geodesics-small-times}
\end{align}
The plot of the function \eqref{chi-of-t-timelike-geodesics-small-times} is given in Fig. \ref{fig:timelike-geodesic-small-time}. 

Although we have shown that both null and timelike geodesics have  no pathological behaviour, it should be noted that  the corresponding velocities blow up at the BB, as it is clear  from Eqs. \eqref{null-geodesic-in-terms-of-t}
and \eqref{timelike-geodesic-in-terms-of-t}.

\subsubsection{Singularity or not? \qm{Mild singularity} or a new kind of singularity}
\label{Sec:mild-singularity}

In the general relativity framework, singularity theorems are based on the criterion that 
timelike and null geodesic completeness are  minimum conditions for a spacetime to be considered singularity-free \cite{Hawking-Ellis,Hawking-Penrose(1970),Wald}. However, these theorems do not prove that singularities of spacetime are necessarily  related to unboundedly large curvature. Indeed,  the  characterization of singularities via the divergent behaviour of the curvature can be inadequate in some situations (e.g., the case of conical singularity). 

Our model is not framed in general relativity, but emerges in the semiclassical limit of the matrix theory. In other words, the cosmic scale factor  occurring in the metric \eqref{eff-metric-FRW} is not a solution of Einstein field equations.   In our analysis,  $a(\eta)$  vanishes at the BB, where, in addition, the  invariants \eqref{kretschmann-invariant}--\eqref{euler-scalar} blow up. Furthermore, the effective metric \eqref{eff-metric-FRW} is zero (and hence  is degenerate) at $\eta=0$. Despite that, no pathological behaviour occur in the analysis of null and timelike geodesics, as we have shown before. Therefore, if we are to make a comparison with the recipes of general relativity, we could say that our FLRW solution features a \qm{mild singularity}\footnote{By \qm{mild singularity} we do not refer to the kind of singularities studied in Refs. \cite{Nojiri2005,Odintsov2022}.} which does not prevent both massless particles and non-comoving observers  from crossing the BB, but introduces  some diverging curvature scalars. Moreover, owing to the degenerate character of the metric \eqref{eff-metric-FRW}, another option is possible, where we could conclude that we are dealing with   a new kind of singularity, which, in principle,  might be present also in general relativity, as discussed by Hawking in Ref. \cite{Hawking1967}.

\subsection{Cosmological observables} \label{Sec:Cosmol-obs}

In this section, we evaluate some cosmological observables of our FLRW geometry. In Sec. \ref{Sec:particle-horizon} we consider the particle horizon, while in Sec. \ref{Sec:lumin-distance} we deal with the luminosity distance. 

\subsubsection{The particle horizon}\label{Sec:particle-horizon}

In terms of the time coordinate, the particle horizon  $\mathscr{D}_{\rm hor}(t)$ at $t >0$ reads as \cite{Weinberg1972}
\begin{align}
    \mathscr{D}_{\rm hor}(t)= a(t) \lim_{t^\star \to + \infty} \int^{t}_{-t^\star} \dfrac{dt^\prime }{\vert a(t^\prime)\vert },
\end{align}
where we note that, for the background under consideration, we are free to choose $-t^\star$
arbitrarily negative. In general relativity settings, this operation would be correct only if the spacetime were singularity-free. However, 
following our discussion of Sec. \ref{Sec:mild-singularity}, it makes sense to  consider such a limit  in 
our model, since we are dealing with a new kind of singularity which cannot be totally explained with the standard tools of Einstein theory. 

Bearing in mind Eqs. \eqref{a-eta} and \eqref{dt-squared}, the particle horizon $\mathscr{D}_{\rm hor}(\eta)$ at $\eta >0$ is 
\begin{align}
\mathscr{D}_{\rm hor}(\eta)&= a(\eta) \lim_{\eta^\star \to + \infty} \int^{\eta}_{-\eta^\star} d\eta^\prime \vert \tanh \eta^\prime \vert 
\nonumber \\
&= \left(R \cosh \eta \sqrt{\sinh \eta}\right) \lim_{\eta^\star \to + \infty}  \left[ \log \left(\cosh \eta^\star\right)+\log \left(\cosh \eta\right)\right] = + \infty,
\end{align}
where $\eta^\star \equiv \eta(t^\star)$. Since $\mathscr{D}_{\rm hor}(\eta)$ diverges, it is possible to receive signals from any coming particle of the spacetime. This can be interpreted as a hint that our model has no need for inflation. 

\subsubsection{Luminosity distance}\label{Sec:lumin-distance}

Let us consider  a light signal  emitted by a comoving source at time $t_e>0$ which is then detected by a comoving observer at time $t_0>t_e$. If we further suppose that the signal moves  keeping  $\theta$ and $\varphi$ constant, the luminosity distance $d_L(t_e;t_0)$ is given by \cite{Weinberg1972}
\begin{align}
   d_L(t_e;t_0) &= r_e \dfrac{a^2(t_0)}{a(t_e)}, 
\end{align}
where the coordinate distance $r_e$ travelled by the signal is expressed by the relation
\begin{align}
    \int_{t_e}^{t_0}\dfrac{dt}{a(t)} = \int_{0}^{r_e}\dfrac{dr}{\sqrt{1+r^2}}= {\rm arcsinh \,}r_e.
\end{align}

In our model, it follows from Eqs. \eqref{r-sinh-chi}--\eqref{dt-squared} that the luminosity distance can be expressed as
\begin{align}
   d_L(\eta_e;\eta_0) &= \dfrac{a^2(\eta_0)}{a(\eta_e)}\sinh\left[\int^{\eta_0}_{\eta_e} d \eta \tanh \eta\right]=\dfrac{a^2(\eta_0)}{a(\eta_e)} \sinh \left[\log\left(\cosh \eta_0\right)-\log\left(\cosh \eta_e\right)\right], 
\end{align}
where $\eta_e \equiv \eta(t_e)>0$ and  $\eta_0 \equiv \eta(t_0)>\eta_e$. It is easy to see that 
\begin{align}
    \lim_{\eta_e \to 0^{+}}  d_L(\eta_e;\eta_0) = + \infty,
\end{align}
which represents an expected result since the cosmic scale factor $a(\eta)$ vanishes at the BB.

\section{The scalar modes on $\cM^{3,1}$} \label{Sec:scalar-modes}

In this second part of the paper, we would like to compute the propagator for a scalar field on the above background, and see if there are any interesting effects due to the BB.
A priori it is not evident how to define the propagator, due to the boundary provided by the BB.
We take as starting point the definition as 2-point function defined by a Gaussian integral in the matrix model (or rather its semi-classical limit).
Schematically,
\begin{align}
\langle\phi(x) \phi(y)\rangle 
= \int dk \, \langle \phi_k(x) \phi_k(y)\rangle, 
\end{align}
where
\begin{align}
 \langle \phi_k(x) \phi_k(y)\rangle
= \frac 1Z \int d\phi \,  \phi_k(x) \phi_k(y) e^{i S[\phi_k]},
\end{align}
$Z$ being the generating functional. The necessary details for this computation  will be provided below.

\subsection{Relevant Operators}

In this section, we introduce some relevant wave operators which will play a crucial role in our forthcoming analysis. 

\subsubsection{The effective d'Alembertian $\Box_G$ }

The metric \eqref{eff-metric-FRW} is  encoded in the  
\qm{matrix} d'Alembertian
\begin{align}
 \Box \ = [T^\mu,[T_\mu,.]]  \sim \ -\{t^\mu,\{t_\mu,.\}\}, 
 \label{Box-def}
\end{align}
which governs the propagation of a scalar fields $\phi$, and is related to the metric d'Alembertian  through
\begin{subequations}
\begin{align}
\Box \  &\sim  |\sinh^{3} \eta| \Box_G, 
\label{Box-def-2}
\\
\Box_G &= -\frac{1}{\sqrt{|G|}}\del_\mu\big(\sqrt{|G|}\, G^{\mu\nu}\del_\nu\big).
\end{align}
\end{subequations}
This can be seen by rewriting the action \eqref{scalar-action-metric} as follows
\begin{align}
S[\phi] = - \int\Omega
\phi \{T^\mu\{T_\mu,\phi\}\} 
= -\int d^4 x\,\sqrt{|G|} \phi \Box_G\phi \ 
\label{scalar-action-Box}
\end{align}
where 
\begin{align}
  \Omega &= \frac{1}{\vert \sinh(\eta)\vert } d^4 x
  = \cosh^3(\eta) d\eta\sinh^2(\chi) d\chi \sin(\theta)d\theta d\varphi
  \label{symplectic-volume-form}
\end{align}
is the $SO(4,1)$-invariant  volume form on $H^4$ in Cartesian and hyperbolic coordinates \eqref{embedding-3d-hyperboloid}, respectively (hereafter, we suppose that $\chi >0$). 
Explicitly, one finds 
\begin{align}
 \Box_G \phi &=  \frac{1}{R^2 \vert \sinh^3 \eta \vert \cosh^3\eta} 
 \del_\eta\big(\cosh^3(\eta)\del_\eta \phi\big)
+ \frac{1}{\vert \sinh \eta\vert}\Delta^{(3)}_\eta \phi,
\nonumber \\
&=\dfrac{1}{R^2} \left(\dfrac{3}{\sinh \eta \vert \sinh \eta \vert \cosh \eta} \partial_\eta +\dfrac{1}{\vert \sinh \eta\vert^3 }\partial^2_\eta\right)\phi + \frac{1}{\vert \sinh \eta \vert}\Delta^{(3)}_\eta \phi,
 \label{G-Box-relation}
\end{align}
(cf. (2.32) in \cite{Steinacker:2019dii})
where  $\Delta^{(3)}_\eta$ is the Laplacian  on the space-like three-dimensional hyperboloid $H^3$
\begin{align}
 \Delta^{(3)}_\eta \phi &= - \frac{1}{R^2\cosh^2\eta} \Biggl[  \left(\dfrac{2}{\tanh \chi}\partial_\chi + \partial^2_\chi\right) + \dfrac{1}{\sinh^2 \chi} \left(\dfrac{1}{\tan \theta}
\partial_\theta + \partial^2_\theta + \dfrac{1}{\sin^2 \theta}\partial^2_\varphi\right) \Biggr]\phi
\nonumber \\
&\equiv - \frac{1}{R^2\cosh^2\eta} \Delta^{(3)}\phi,
  \label{Delta-3-H}
\end{align}
where we have defined
\begin{align}
    \Delta^{(3)} \phi \equiv \Biggl[  \left(\dfrac{2}{\tanh \chi}\partial_\chi + \partial^2_\chi\right) + \dfrac{1}{\sinh^2 \chi} \left(\dfrac{1}{\tan \theta}
\partial_\theta + \partial^2_\theta + \dfrac{1}{\sin^2 \theta}\partial^2_\varphi\right) \Biggr] \phi. 
\label{Delta-3-without-eta}
\end{align}
We note that to derive Eq. \eqref{Delta-3-H} we have exploited that the metric on the space-like $H^3$  is
\begin{equation}
    ds^2\vert_{H^3}= R^2 \cosh^2 (\eta) d\Sigma^2.
\end{equation}

\subsubsection{The Laplacian operator $\Delta_{\mathscr{G}}$  on $H^4$}

The induced metric on the four-dimensional hyperboloid $H^4$ is (cf. Eq. \eqref{embedding-4d-hyperboloid})
\begin{align}
    ds^2\vert_{H^4}= \mathscr{G}_{\mu \nu}dx^\mu dx^\nu= R^2 d\eta^2 + R^2 \cosh^2 (\eta) d\Sigma^2,
    \label{induced-metric-on-H4eta}
\end{align}
where the length element $d\Sigma^2$ on a spatial standard three-dimensional hyperboloid $H^3$ has been given in Eq. \eqref{dSigma2}. By means of the metric \eqref{induced-metric-on-H4eta}, the Laplacian operator $ \Delta_{\mathscr{G}}$ on a generic function  $\phi=\phi(\eta,\chi,\theta,\varphi)$  reads as
\begin{align}
    \Delta_{\mathscr{G}} \phi &= \dfrac{1}{\sqrt{\mathscr{G}}}\partial_\mu \left(\sqrt{\mathscr{G}}\mathscr{G}^{\mu \nu} \partial_\nu \phi\right)=\dfrac{1}{R^2}\Biggl\{ 3 \tanh (\eta) \partial_\eta  + \partial^2_\eta  +\dfrac{1}{\cosh^2 (\eta)} \Biggl[ \dfrac{2}{\tanh(\chi)}\partial_\chi  +\partial^2_\chi  
    \nonumber \\
   & +\dfrac{1}{\sinh^2(\chi)} \left(\dfrac{1}{\tan(\theta)}\partial_\theta +\partial^2_\theta  +\dfrac{1}{\sin^2(\theta) } \partial^2_\varphi \right) \Biggr]\Biggr\}\phi,
   \label{Laplacian-on-H4}
\end{align}
where $\sqrt{\mathscr{G}}=R^4 \cosh^3 (\eta) \sinh^2 (\chi) \sin (\theta)$.

\subsubsection{Relations between $\Delta_{\mathscr{G}}$ and $\Delta^{(3)}_\eta$ and $\Box_G$ and $\Delta^{(3)}_\eta$}

The Laplacian $\Delta_{\mathscr{G}}$ on the four-dimensional hyperboloid $H^4$ and the Laplacian $\Delta^{(3)}_\eta$ on the spacelike three-dimensional hyperboloid $H^3$ are related by (cf. Eqs. \eqref{Delta-3-H} and \eqref{Laplacian-on-H4})
\begin{align}
    \Delta_{\mathscr{G}} = \dfrac{1}{R^2} \left(3 \tanh (\eta) \partial_\eta + \partial^2_\eta\right) - \Delta^{(3)}_\eta.
\label{relation-Delta-G-and-Delta-3}    
\end{align}
On the other hand, the effective d'Alembertian $\Box_G$ is related to $\Delta^{(3)}_\eta$ as (cf. Eq. \eqref{G-Box-relation})
\begin{align}
\vert \sinh^3 \eta \vert    \Box_G = \dfrac{1}{R^2} \left(3 \tanh (\eta) \partial_\eta + \partial^2_\eta\right) +\left(\sinh^2 \eta \right) \Delta^{(3)}_\eta.
\label{relation-Box-G-and-Delta-3}    
\end{align}

Last, we observe that the effective d'Alembertian $\Box_G$ is related to the Laplacian $\Delta_\mathscr{G}$ on $H^4$  as follows (cf. Eqs. \eqref{G-Box-relation} and \eqref{Laplacian-on-H4})
\begin{align}
\label{relation-Box-G-Delta-G}
   R^2 \vert \sinh^3 \eta \vert \Box_G = R^2 \Delta_\mathscr{G} - \left(\dfrac{2}{\tanh \chi}\partial_\chi+ \partial^2_\chi\right) -\dfrac{1}{\sinh^2 \chi}\left(\dfrac{1}{\tan \theta}\partial_\theta + \partial_\theta^2 + \dfrac{1}{\sin^2 \theta}\partial^2_\varphi\right).
\end{align}

An inspection of Eqs.  \eqref{relation-Delta-G-and-Delta-3}--\eqref{relation-Box-G-Delta-G} reveals that the eigenfunctions of  the operators $\Delta_{\mathscr{G}}$, $\Delta^{(3)}$, and  $\Box$  share the same building blocks. The eigenfunctions of  d'Alembertian operator $\Box$ are derived in the next section, while those  of $\Delta_{\mathscr{G}}$  and $\Delta^{(3)}$ can be found in Appendix \ref{Sec:Eigenfunctions-of-Laplacian-operators}.

\subsection{Eigenfunctions of the  d'Alembertian operator $\Box$}\label{Sec:Eigenfunctions-of-Box}

It is convenient to work out the eigenfunctions of the operator $\Box$ instead of $\Box_G$ for at least  two reasons. First of all, the d'Alembertian $\Box$ is self-adjoint with respect to the symplectic volume form \eqref{symplectic-volume-form}; secondly, the operator $\Box$ admits handier eigenfunctions than $\Box_G$.

Bearing in mind Eqs. \eqref{Box-def-2}, \eqref{G-Box-relation}, and  \eqref{Delta-3-H}, the d'Alembertian operator $\Box$ reads as
\begin{align}
    \Box \phi &= \dfrac{1}{R^2}\Biggl[ 3 \tanh (\eta) \partial_\eta  + \partial^2_\eta  -\tanh^2 \eta \left( \dfrac{2}{\tanh \chi}\partial_\chi  +\partial^2_\chi  \right)
    \nonumber \\
   & -\dfrac{\tanh^2 \eta}{\sinh^2 \chi} \left(\dfrac{1}{\tan \theta}\partial_\theta +\partial^2_\theta  +\dfrac{1}{\sin^2 \theta } \partial^2_\varphi \right) \Biggr]\phi.
   \label{Box-expression}
\end{align}

The eigenfunctions of the  d'Alembertian operator are defined by the equation
\begin{align}
    \Box \phi = \lambda \phi,
    \label{eigenvalue-problem-BoxG}
\end{align}
whose resolution can be tackled via the separation \emph{ansatz}
\begin{subequations}
\begin{align}
    \phi(\eta,\chi,\theta,\varphi) &= \tilde{\phi}(\eta,\chi)Y^m_l(\theta,\varphi),
   \\
    \tilde{\phi}(\eta,\chi)&=f(\eta)g(\chi),
\end{align}
\label{separation-ansatz-BoxG}
\end{subequations}
where the spherical harmonic functions $Y^m_l(\theta,\varphi)$  of degree $l$ and order $m$ (with $l \geq \vert m \vert$)  satisfy the well-known property \cite{Jackson1998}
\begin{align}
    \left(\dfrac{1}{\tan\theta}\partial_\theta +\partial^2_\theta  +\dfrac{1}{\sin^2\theta } \partial^2_\varphi \right)Y^m_l(\theta,\varphi)=-l(l+1)Y^m_l(\theta,\varphi).
    \label{spherical-harmonics-property}
\end{align}

Bearing in mind Eqs. \eqref{separation-ansatz-BoxG} and \eqref{spherical-harmonics-property}, the eigenvalue problem \eqref{eigenvalue-problem-BoxG} gives
\begin{align}
    \dfrac{1}{f}\left(3 \tanh (\eta) \partial_\eta  + \partial^2_\eta \right)f -\dfrac{\tanh^2 \eta}{g} \left(\dfrac{2}{\tanh\chi}\partial_\chi  +\partial^2_\chi\right) g +\dfrac{l(l+1)\tanh^2 \eta}{ \sinh^2 \chi} = \lambda R^2, 
\end{align}
where we have divided both sides by  $\tilde{\phi}(\eta,\chi)Y^m_l(\theta,\varphi)$. After having performed some manipulations, the above equation  can be solved through the method of separation of variables, yielding the following two ordinary differential equations:
\begin{subequations}
\label{ode-eta&chi-BoxG-operator}
\begin{align}
\left( \partial^2_\eta + 3 \tanh (\eta) \partial_\eta  -\beta  \tanh^2 \eta-\lambda R^2 \right)f(\eta)&=0,
\label{ode-eta-BoxG}
\\
\left(\partial^2_\chi +\dfrac{2}{\tanh \chi} \partial_\chi -\dfrac{l(l+1)}{\sinh^2 \chi}  - \beta\right)g(\chi) &=0,
\label{ode-chi-BoxG}
\end{align}
\end{subequations}
$\beta$ being a real-valued constant.

The solution of Eq. \eqref{ode-eta&chi-BoxG-operator} and the eigenfunctions of the d'Alembertian  operator \eqref{Box-expression} will be provided in the next sections.

\subsubsection{The time-like equation}

It is instructive to work out the details of the solution of  Eq. \eqref{ode-eta-BoxG}. If we introduce the variable
\begin{align}
    w=\tanh \eta \in (-1,1),
    \label{w-variable}
\end{align}
then the derivative operators read as
\begin{align}
    \partial_\eta &= (1-w^2)\partial_w, 
    \nonumber \\
    \partial^2_\eta &= -2w(1-w^2)\partial_w + (1-w^2)^2 \partial^2_w,
\end{align}
and hence Eq. \eqref{ode-eta-BoxG} becomes
\begin{align}
    \left[(1-w^2) \partial^2_w + w \partial_w - \beta\dfrac{w^2}{1-w^2} -\dfrac{\lambda R^2}{1-w^2} \right]f(w)=0.
\end{align}
If we write
\begin{align}
    f(w)=(1-w^2)^{3/4}h(w),
    \label{f(w)-and-h(w)}
\end{align}
then we end up with the general Legendre equation \cite{Abramowitz-Stegun(1964),Magnus1966}
\begin{align}
    (1-w^2)\partial^2_w h -2w\partial_w h + \left[\left(\frac{3}{4}+\beta\right) -\left(\frac{\frac{9}{4} +\beta+\lambda R^2}{1-w^2}\right)\right]h=0,
\end{align}
which can be  solved via the associated Legendre functions of the first and second kind $\mathsf{P}^\mu_\nu(w)$ and $\mathsf{Q}^\mu_\nu(w)$, respectively, having   degree $\nu$ and order $\mu$ given by
\begin{subequations}
\begin{align}
\nu &= \dfrac{1}{2}\left(2\sqrt{1+\beta}-1\right),
\label{degree-nu-BoxG}
\\
\mu &= \dfrac{1}{2}\sqrt{9+4\beta+4\lambda R^2}.
\label{order-mu-BoxG}
\end{align}
\label{degree-nu-&-order-mu-BoxG}
\end{subequations}
Therefore, bearing in mind Eqs. \eqref{w-variable} and \eqref{f(w)-and-h(w)},  the solution of Eq.  \eqref{ode-eta-BoxG} in terms of the variable $\eta$ is
\begin{align}
    f(\eta)= (1-\tanh^2 \eta)^{3/4}\left[ c_1 \mathsf{P}^\mu_\nu(\tanh \eta) +c_2  \mathsf{Q}^\mu_\nu(\tanh \eta)\right],
\label{f-eta-solution-BoxG}    
\end{align}
$c_1$ and $c_2$ being  integration constants. 
Here $\mathsf{Q}^{\mu}_{\nu}$ can equivalently be replaced by $\mathsf{P}^{- \mu}_{\nu}$, which will be done in the following.

\subsubsection{The radial equation}

The solution of \eqref{ode-chi-BoxG} can be written in terms of the Gaussian or ordinary hypergeometric function \cite{Abramowitz-Stegun(1964)} as
\begin{align}
    g(\chi) &= \dfrac{\left(1-\tanh^2 \chi\right)^{(1/2)(1-a+b+c)}}{\left(\tanh\chi\right)^{3/2}} \Biggl[ c_3 \left(\tanh^2 \chi\right)^{a/2} {}_{2}\mathrm{F}_{1}\left(b,c;a;\tanh^2 \chi\right) 
\nonumber \\    
    &+ c_4 \left(-1\right)^{1-a}\left(\tanh^2 \chi\right)^{1-a/2} {}_{2}\mathrm{F}_{1}\left(1-a+b,1-a+c;2-a;\tanh^2 \chi\right)\Biggr],
\label{g-chi-solution-BoxG}
\end{align}
where $c_3$ and $c_4$ are    integration constants and
\begin{align}
a & \equiv \dfrac{1}{2} \left(3+2l\right),
\nonumber \\
b & \equiv \dfrac{1}{2}  \left(1+l+\sqrt{1+\beta}\right),
\nonumber \\
c & \equiv \dfrac{1}{2}  \left(2+l+\sqrt{1+\beta}\right) = b + \dfrac{1}{2}.
\label{a-b-c-BoxG}
\end{align}

It is possible to write Eqs. \eqref{ode-chi-BoxG} and \eqref{g-chi-solution-BoxG} in a more convenient form. Indeed, if we introduce the variable (recall that we are considering $\chi >0$)
\begin{equation}
    y=\coth \chi \in  (1,+\infty), 
\label{y-variable}
\end{equation}
then we have
\begin{align}
\partial_\chi &= (1-y^2)\partial_y,
\nonumber \\
\partial^2_\chi &= (1-y^2)^2\partial^2_y -2y(1-y^2)\partial_y,    
\label{partial-chi-and-partial-y}
\end{align}
and hence Eq. \eqref{ode-chi-BoxG} becomes
\begin{equation}
    (1-y^2)\partial^2_y g(y)+ \left[l(l+1)-\dfrac{\beta}{1-y^2}\right]g(y)=0.
\end{equation}
The solution of the above equation reads as
\begin{align}
g(\chi)=\sqrt{\coth^2 \chi -1}  \left[c_3 \mathcal{P}^{\tilde{\mu}}_l(\coth \chi) +c_4  \mathcal{Q}^{\tilde{\mu}}_l(\coth \chi) \right],  
\label{g-chi-solution-Legendre-BoxG}
\end{align}
where we have exploited Eq. \eqref{y-variable} and
\begin{align} \label{tilde-mu}
    \tilde{\mu} \equiv  \sqrt{1+\beta}.
\end{align}
The physically meaningful solutions can be identified by considering the boundary conditions. Recall that $\chi\in[0,\infty)$ plays the role of a radial variable.
In the limit $\chi \to 0$, we have 
\eqref{asymptpotic-Legendre-Q-infinity}
\begin{align}
   \mathcal{Q}^{\tilde \mu}_l(\coth\chi) \sim (\coth\chi)^{-l-1} \sim \chi^{l+1} 
\end{align}
since $\coth \chi \sim \chi^{-1}$, while 
 $\mathcal{P}^{\tilde \mu}_l(\coth \chi)$ is divergent for $l\geq 0$,
and finite for $l=0$.
Therefore to have non-singular
solutions, we should choose  the $\mathcal{Q}^{\tilde\mu}_l$ solutions and reject the $\mathcal{P}^{\tilde\mu}_l$. 

\subsubsection{The eigenmodes on the spacetime}

Thanks to the solutions \eqref{f-eta-solution-BoxG} and \eqref{g-chi-solution-Legendre-BoxG}, the eigenfunctions \eqref{separation-ansatz-BoxG} of the d'Alembertian  operator \eqref{Box-expression} are known.   Moreover, Eq. \eqref{order-mu-BoxG} permits  obtaining the explicit expression of the eigenvalue $\lambda$, which reads as
\begin{align}
 \lambda= \dfrac{ \mu^2-\beta -\frac{9}{4}}{R^2}.  
 \label{lambda-express-1} 
\end{align}
In order to have oscillatory (square-integrable) solutions, we should assume that the order \eqref{order-mu-BoxG} of the solution  \eqref{f-eta-solution-BoxG} is  purely imaginary. Therefore, we set
\begin{align}
\mu = \pm is,
\label{mu-equals-i-s}
\end{align}
where
\begin{align}
s= \vert \mu \vert =\sqrt{-\left(\dfrac{9}{4}+\beta+\lambda R^2\right)} \ >0.
\label{s-equals-vert-mu}    
\end{align}
Similarly, the order \eqref{tilde-mu} of the solution \eqref{g-chi-solution-Legendre-BoxG} should be purely imaginary, i.e.
\begin{align}
     \tilde{\mu}=  i q \ .
\end{align}
for $q\in\R$. This leads to 
\begin{align}
    q^2 = -\left(1+\beta\right)  >0 \ ,
\label{q-equals-vert-tilde-mu}    
\end{align}
which implies that 
\begin{align}
    \beta < -1. 
    \label{beta-less-than-1}
\end{align}
From Eqs. \eqref{s-equals-vert-mu} and \eqref{q-equals-vert-tilde-mu}, the eigenvalue  \eqref{lambda-express-1} can be written equivalently as
\begin{align}
   \lambda= \dfrac{ q^2-s^2 -\frac{5}{4}}{R^2}.  
 \label{lambda-expression}    
\end{align}

As a consequence of Eq. \eqref{beta-less-than-1}, we see that the degree \eqref{degree-nu-BoxG} of the solution \eqref{f-eta-solution-BoxG} is complex and reads as 
\begin{align}
    \nu= -\dfrac{1}{2} \pm iq \ .
\end{align}
In order not to overcount equivalent solutions, we observe that the $\mathsf{P}^\mu_\nu(\tanh \eta)$ coincide for the two choices of $ \nu= -\dfrac{1}{2} \pm iq$. We can therefore make the convention that 
\begin{align}
\nu= -\dfrac{1}{2} + i|q|.
\label{degree-nu-with-plus}
\end{align}
Collecting Eqs. \eqref{mu-equals-i-s}--\eqref{q-equals-vert-tilde-mu} jointly with  Eqs. \eqref{f-eta-solution-BoxG} and \eqref{g-chi-solution-Legendre-BoxG},  the eigenmodes \eqref{separation-ansatz-BoxG} of the d'Alembertian  operator \eqref{Box-expression} having the appropriate boundary conditions are
\begin{align}
\boxed{\ 
    \Upsilon^{s_\pm,q}_{l,m}\left(\eta,\chi,\theta,\varphi\right):=\dfrac{1}{\sqrt{\cosh^3 \eta} \sinh \chi }\mathsf{P}^{\pm is}_\nu \left(\tanh \eta \right) \mathcal{Q}^{iq}_l \left(\coth \chi\right) Y^m_l(\theta,\varphi), \quad q \in\R$, $s>0
    \ }
\label{Upsilon}    
\end{align}
recalling that $\chi>0$. The above-defined eigenmodes are regular functions in the limit $\chi \to 0$.
Indeed,  due to the relation \eq{asymptpotic-Legendre-Q-infinity}, we have
\begin{align}
    \Upsilon^{s_\pm,q}_{l,m}(\chi) \sim \chi^{l}, \qquad \mbox{as} \quad \chi \to 0 \ .
\label{asymptotic-chi-1}
\end{align}
Now consider the limit $\chi \to +\infty$, which means that 
$\coth \chi \to 1^+$. The functions $\mathcal{Q}^{iq}_l(x)$ are oscillatory but bounded when $x\to 1^+$ as  it can be inferred from Eq. \eqref{asymptpotic-Legendre-Q-1plus}. In this way, we obtain
\begin{align}
   \Upsilon^{s_\pm,q}_{l,m}\left(\chi\right) \sim  \frac{e^{i\chi q}}{\sinh(\chi)},  
   \qquad \mbox{as} \quad \chi \to +\infty. 
\end{align}
Therefore, the eigenfunctions  $\Upsilon^{s_\pm,q}_{l,m}\left(\eta,\chi,\theta,\varphi\right)$
are square-integrable oscillatory functions on $\cM^{3,1}$, as it should be.
Moreover, they are continuous across the BB
i.e. at $\eta = 0$. 

\subsubsection{Flat regime}

Now consider the following 
\qm{flat} regime\footnote{We will consider this regime also in the computation of the propagator for $\eta \to 0$, where the geometry is strictly speaking no longer flat.}  
\begin{align}
\mbox{FR}:\quad  \chi  < 1, \qquad q \gg l,
\label{regime-A}
\end{align}
where \qm{FR} stands for \qm{flat regime}, and $q$ will be a typical momentum.
Then the $\mathcal{Q}^{iq}_l$ reduces to the spherical Bessel functions, 
\begin{align}
  {\rm Re}\left[\frac{\mathcal{Q}^{iq}_l(\coth \chi)}{\sinh \chi}  \frac{q^l}{e^{-\pi q} \Gamma(i q + l + 1)} \right]
  \overunderset{\chi <1}{q \gg l}{\sim}\, j_l(q\chi).
  \label{Legendre-bessel-flat}
\end{align}
This should be expected, since the eigenfunctions \eqref{Upsilon} should reduce to the standard ones on $\R^{3,1}$. For $\chi \to 0$, the relation \eqref{Legendre-bessel-flat} is guaranteed by the standard expansion formulas for $\mathcal{Q}^{iq}_l(\coth \chi)$ and $ j_l(q\chi)$\footnote{The expansion  of $\mathcal{Q}^{iq}_l(\coth \chi)$ when $\chi \to 0$ can be read off from Eq. \eqref{asymptpotic-Legendre-Q-infinity}; for  $ j_l(q\chi)$ we refer the reader to Ref. \cite{Jackson1998}.}; here, we have verified that
it  works very well in the range $\chi\in(0,1)$ for $q\gg l$. Moreover, it also holds that
\begin{align}
     {\rm Im}\left[\frac{\mathcal{Q}^{iq}_l(\coth \chi)}{\sinh \chi}  \frac{q^l}{e^{-\pi q} \Gamma(i q + l + 1)} \right] \overset{\chi <1}{\sim} 0,
\end{align}
and hence we can conclude that
\begin{align}
\frac{\mathcal{Q}^{iq}_l(\coth \chi)}{\sinh \chi}  \frac{q^l}{e^{-\pi q} \Gamma(i q + l + 1)} \,
 \ \overunderset{\chi <1}{q \gg l}{\sim}\, \ j_l(q\chi).
 \label{Legendre-bessel-flat-complete}
\end{align}
Due to the  relation \cite{Olver1997}
\begin{align}
\mathsf{P}^\mu_\nu(x) &\overset{x \to 1^{-}}{\sim} \dfrac{1}{\Gamma(1-\mu)}  \left(\dfrac{1+x}{1-x}\right)^{\mu/2},
\end{align}
we easily obtain
\begin{align}
    \mathsf{P}^{\pm is}_\nu \left(\tanh \eta \right) &\overset{\eta \to +\infty}{\sim} \dfrac{1}{\Gamma(1 \mp is)} e^{\pm i \eta s}.
\label{Legendre-P-eta-plus-infinity}    
\end{align}
Thus, bearing in mind Eqs.   \eqref{Legendre-bessel-flat-complete} and \eqref{Legendre-P-eta-plus-infinity}, the eigenmodes \eqref{Upsilon} for large times (i.e., $\eta \to + \infty$) and in the flat regime \eqref{regime-A} become
\begin{align}
\Upsilon^{s_{\pm},q}_{l,m}(\eta,\chi,\theta,\varphi)&\overunderset{\eta \to + \infty}{{\rm FR}}{\sim}\, \dfrac{1}{\sqrt{\cosh^3 \eta}}\dfrac{e^{-\pi q} j_l\left(q \chi\right) \Gamma(iq+l+1)}{ q^l} \dfrac{e^{\pm i \eta s}}{\Gamma(1\mp is)}Y^{m}_l(\theta,\varphi).
\label{flate-regime-Upsilon}
\end{align}

  \subsection{Orthogonality relations}

The orthogonality relations for the eigenfunctions \eqref{Upsilon}  of the d'Alembertian  operator \eqref{Box-expression} are written via the $SO(4,1)$-invariant inner product as 
\begin{align}
& \langle \Upsilon^{s^\prime_{+},q^\prime}_{l^\prime,m^\prime}, \Upsilon^{s_{+},q}_{l,m}\rangle  \left(\eta,\chi,\theta,\varphi\right):= \int \Omega \left[\Upsilon^{s^\prime_{+},q^\prime}_{l^\prime,m^\prime}\left(\eta,\chi,\theta,\varphi\right)\right]^* \Bigl[\Upsilon^{s_{+},q}_{l,m} \left(\eta,\chi,\theta,\varphi\right)\Bigr],
\label{orthogonality-relations-1}    
\end{align}
 the  symplectic volume form $\Omega$ being given in  Eq. \eqref{symplectic-volume-form}. Explicitly, we obtain
 \begin{align}
& \langle \Upsilon^{s^\prime_{+},q^\prime}_{l^\prime,m^\prime}, \Upsilon^{s_{+},q}_{l,m}\rangle  = \int_0^{2 \pi}d \varphi \int_0^\pi d \theta \sin \theta \left[Y^{m^\prime}_{l^\prime}\left(\theta,\varphi\right)\right]^* \Bigl[Y^{m}_{l}\left(\theta,\varphi\right)\Bigr]
\nonumber \\      
& \times \int^{+\infty}_{-\infty} d\eta \left[\mathsf{P}^{ is^\prime}_{\nu^\prime}\left(\tanh \eta \right)\right]^* \Bigl[ \mathsf{P}^{ is}_\nu\left(\tanh \eta \right)\Bigr] \int^{+\infty}_{0} d \chi  \left[\mathcal{Q}^{iq^\prime}_{l^\prime}\left(\coth \chi \right)\right]^* \Bigl[\mathcal{Q}^{ iq}_l\left(\coth \chi \right)\Bigr]
\nonumber \\
&= \delta_{l l^\prime} \delta_{m m^\prime} \int^{+\infty}_{-\infty} d\eta \left[\mathsf{P}^{ is^\prime}_{\nu^\prime}\left(\tanh \eta \right)\right]^* \Bigl[ \mathsf{P}^{ is}_\nu\left(\tanh \eta \right)\Bigr] \int^{+\infty}_{0} d \chi  \left[\mathcal{Q}^{iq^\prime}_{l}\left(\coth \chi \right)\right]^* \Bigl[\mathcal{Q}^{ iq}_l\left(\coth \chi \right)\Bigr],
\label{orthogonality-relations-2}    
 \end{align}
where we have exploited the orthogonality of the spherical harmonics \cite{Jackson1998} and the ensuing Kronecker delta factor $\delta_{l l^\prime}$. We first consider the integral involving the $\chi$ variable.  If we define $\xi = \tanh \chi $, then we obtain
\begin{align}
\int^{+\infty}_{0} d \chi  \left[\mathcal{Q}^{iq^\prime}_{l}\left(\coth \chi \right)\right]^* \Bigl[\mathcal{Q}^{ iq}_l\left(\coth \chi \right)\Bigr] =\int_0^1 \dfrac{d \xi}{1-\xi^2}\left[\mathcal{Q}^{iq^\prime}_{l} (1/\xi)\right]^* \mathcal{Q}^{ iq}_l(1/\xi),
\end{align}
which, via the substitution $x = 1/\xi$, yields
\begin{align}
   \int_0^1 \dfrac{d \xi}{1-\xi^2}\left[\mathcal{Q}^{iq^\prime}_{l} (1/\xi)\right]^* \mathcal{Q}^{ iq}_l(1/\xi) &=-\int_1^{+\infty} \dfrac{dx}{1-x^2} \left[ \mathcal{Q}^{iq^\prime}_{l} (x)\right]^* \mathcal{Q}^{ iq}_l(x).
\end{align}
The above integral can be worked out with the help of the results of  Appendix \ref{Sec:Appendix-orthogonality}, where we have shown that 
\begin{align}
\int_1^{+\infty} \dfrac{dx}{1-x^2} \mathcal{Q}^{iq^\prime}_{l} (x) \mathcal{Q}^{ iq}_l(x)=-\dfrac{\left(\pi/2\right)^2}{q \sinh \left(\pi q\right)} \delta(q+q^\prime), \qquad (l>-3).
\label{integral-result-1}
\end{align}
The above relation leads to
\begin{align}
-\int_1^{+\infty} \dfrac{dx}{1-x^2} \mathcal{Q}^{-iq^\prime}_{l} (x) \mathcal{Q}^{ iq}_l(x)=\dfrac{\left(\pi/2\right)^2}{q \sinh \left(\pi q\right)} \delta(q-q^\prime),
\end{align}
which in turn means that, for any real $q,q^\prime$, 
\begin{align}
\int^{+\infty}_{0} d \chi  \left[\mathcal{Q}^{iq^\prime}_{l}\left(\coth \chi \right)\right]^* \Bigl[\mathcal{Q}^{ iq}_l\left(\coth \chi \right)\Bigr]=  \dfrac{\left(\pi/2\right)^2 e^{-2 \pi q}}{q \sinh \left(\pi q\right)} \delta(q-q^\prime),
\label{orth-Q-iq-l-chi}
\end{align} 
where we have exploited (see Eqs. \eqref{Olver-associated-Legendre-associated}--\eqref{star-of-Q-iq-l})
\begin{align}
\left[\mathcal{Q}^{iq^\prime}_l(x)\right]^* = e^{-2 \pi q^\prime} \mathcal{Q}^{-iq^\prime}_{l}(x).
\end{align}
By virtue of the   identity \eqref{orth-Q-iq-l-chi}, Eq. \eqref{orthogonality-relations-2} gives 
\begin{align}
\langle \Upsilon^{s^\prime_{+},q^\prime}_{l^\prime,m^\prime}, \Upsilon^{s_{+},q}_{l,m}\rangle &= \dfrac{e^{-2 \pi q}\left(\pi/2\right)^2}{q \sinh \left(\pi q\right)}  \delta_{l l^\prime} \delta_{m m^\prime} \delta(q-q^\prime) \int^{+\infty}_{-\infty} d\eta \left[\mathsf{P}^{ is^\prime}_{\nu^\prime}\left(\tanh \eta \right)\right]^* \Bigl[ \mathsf{P}^{ is}_\nu\left(\tanh \eta \right)\Bigr] 
\nonumber \\
&=\dfrac{e^{-2 \pi q}\left(\pi/2\right)^2}{q \sinh \left(\pi q\right)}  \delta_{l l^\prime} \delta_{m m^\prime} \delta(q-q^\prime) \int^{+\infty}_{-\infty} d\eta \left[\mathsf{P}^{ is^\prime}_{\nu}\left(\tanh \eta \right)\right]^* \Bigl[ \mathsf{P}^{ is}_\nu\left(\tanh \eta \right)\Bigr].
\label{orthogonality-relations-3}
\end{align}
We recall that the  expression of the associated Legendre function of the first kind with a generic degree $\nu$  and  argument $x$ lying in the interval $(-1,1)$ is \cite{Abramowitz-Stegun(1964),Olver1997}
\begin{align}
 \mathsf{P}^{ is^\prime}_\nu\left(x \right)&= \dfrac{1}{\Gamma(1-is^\prime)}\left(\frac{1+x}{1-x}\right)^{is^\prime/2}
{}_{2}\mathrm{F}_{1}\left(\nu+1,-\nu;1-is^\prime;\tfrac{1}{2}-\tfrac{1}{2}x\right),  \qquad (-1<x<1),   
\label{Legendre-P-hypergeometric-x-minus1-1}
\end{align}
${}_{2}\mathrm{F}_{1}\left(a,b;c;x\right)$ being, like before,  the Gaussian (or ordinary) hypergeometric function and $\Gamma(x)$  the gamma function. In our case, the degree $\nu$ of $\mathsf{P}^{ is^\prime}_{\nu}\left(\tanh \eta \right)$ assumes the form  given in Eq. \eqref{degree-nu-with-plus}. For this  choice of $\nu$, we find
\begin{align}
\left[\mathsf{P}^{ is^\prime}_{\nu}\left(\tanh \eta \right)\right]^* = \mathsf{P}^{- is^\prime}_{\nu}\left(\tanh \eta \right),  
\end{align}
where we have exploited  Eq. \eqref{Legendre-P-hypergeometric-x-minus1-1} jointly with the property  ${}_{2}\mathrm{F}_{1}\left(a,b;c;x\right)={}_{2}\mathrm{F}_{1}\left(b,a;c;x\right)$.
With this result in mind, it follows from Eq. \eqref{orthogonality-relations-3} that
\begin{align}
\langle \Upsilon^{s^\prime_{+},q^\prime}_{l^\prime,m^\prime}, \Upsilon^{s_{+},q}_{l,m}\rangle &= \dfrac{e^{-2 \pi q}\left(\pi/2\right)^2}{q \sinh \left(\pi q\right)}  \delta_{l l^\prime} \delta_{m m^\prime} \delta(q-q^\prime) \int^{+\infty}_{-\infty} d\eta \, \mathsf{P}^{-is^\prime}_{\nu}\left(\tanh \eta \right)  \mathsf{P}^{ is}_\nu\left(\tanh \eta \right)
\nonumber \\
&=\dfrac{e^{-2 \pi q}\left(\pi/2\right)^2}{q \sinh \left(\pi q\right)}  \delta_{l l^\prime} \delta_{m m^\prime} \delta(q-q^\prime) \int^{1}_{-1} \dfrac{dy}{1-y^2} \mathsf{P}^{-is^\prime}_{\nu}\left(y \right)  \mathsf{P}^{ is}_\nu\left(y \right),
\label{orthogonality-relations-4}
\end{align}
where we have employed the substitution $y=\tanh \eta$. This last integral can be calculated with the help of following result:
\begin{align}
\int^1_{-1} \dfrac{dy}{1-y^2}\mathsf{P}^{is^\prime}_\nu(y) \mathsf{P}^{is}_\nu(y) 
&= -\dfrac{2 \pi \sin(\pi \nu)}{s \sinh(\pi s)}\dfrac{1}{\Gamma(1+\nu-is)\Gamma(-\nu-is)}\delta(s-s^\prime)
\nonumber \\
&+2\left[\dfrac{\sinh(\pi s)}{s}+\dfrac{\sin^2(\pi \nu)}{s \sinh(\pi s)}\right]\delta(s+s^\prime),
\label{integral-result-2}
\end{align}
which has been proved in Ref. \cite{Bielski2013} (see Appendix \ref{Sec:Appendix-orthogonality} for further details). Therefore, bearing in mind the above formula along with Eq. \eqref{degree-nu-with-plus}, from Eq. \eqref{orthogonality-relations-4} we finally obtain the sought-after orthogonality relations
\begin{align}
\langle \Upsilon^{s^\prime_{+},q^\prime}_{l^\prime,m^\prime}, \Upsilon^{s_{+},q}_{l,m}\rangle &= \int \Omega \left[\Upsilon^{s^\prime_{+},q^\prime}_{l^\prime,m^\prime}\right]^* \Bigl[\Upsilon^{s_{+},q}_{l,m} \Bigr]
\nonumber \\
&=\dfrac{e^{-2 \pi q}\left(\pi/2\right)^2}{q \sinh \left(\pi q\right)}  \delta_{l l^\prime} \delta_{m m^\prime} \delta(q-q^\prime) \Biggl[a(q,s)\delta(s+s^\prime) + b(q,s)\delta(s-s^\prime)\Biggr],
\label{orthogonality-relations-final}
\end{align}
where 
\begin{subequations}
\label{a-q-s-and-b-q-s}
\begin{align}
a(q,s) &= \dfrac{2 \pi \cosh(\pi q)}{s \sinh(\pi s)}\dfrac{1}{\Gamma\left(iq-is+1/2\right)\Gamma(-iq-is+1/2)}=a(q,-s)^*,
\\
b(q,s) &= \dfrac{2 \sinh(\pi s)}{s}\left[1+\dfrac{\cosh^2(\pi q)}{\sinh^2(\pi s)}\right]=b(q,-s).
\end{align}
\end{subequations}

\subsection{Completeness relation for the $\mathcal{Q}^{iq}_{\nu}$}
\label{sec:completeness-Q}

As a consequence of \eqref{orth-Q-iq-l-chi},
we obtain the following completeness relation for the $\mathcal{Q}^{iq}_{\nu}(\coth \chi)$ in the interval $(1,+\infty)$: 
\begin{align}
    \int_{-\infty}^{+\infty} dq\,
     \dfrac{q \sinh \left(\pi q\right)}{e^{-2 \pi q}\left(\pi/2\right)^2}\,
    \mathcal{Q}^{iq}_{l}(\coth \chi^\prime)
    \left[\mathcal{Q}^{iq}_{l}(\coth \chi)\right]^*
    = \d(\chi-\chi^\prime) \ .
    \label{complete-Q-iq-l-chi}
\end{align}
To prove this, it suffices to multiply this equation with  $\mathcal{Q}^{iq^\prime}_{l}(\coth \chi)$ and integrate over $\int_0^{+\infty} d\chi$ using \eqref{orth-Q-iq-l-chi}.
This completeness relation allows to represent any (square-integrable) function $\phi(\chi)$ defined in the interval $(0,+\infty)$ as superposition of $ \mathcal{Q}^{iq}_{l}$ modes,
\begin{align}
    \phi(\chi) = \int_{-\infty}^{+\infty} dq\,
    \dfrac{q \sinh \left(\pi q\right)}{e^{-2 \pi q}\left(\pi/2\right)^2}\,
    \hat\phi(q)\, \mathcal{Q}^{iq}_{l}(\coth \chi),
\end{align}
where $\hat\phi(q)$ is given by 
\begin{align}
    \hat\phi(q) = \int_0^{+\infty} d\chi\,
    \phi(\chi) \left[\mathcal{Q}^{iq}_{l}(\coth \chi)\right]^*.
\end{align}
The factor $\dfrac{q \sinh \left(\pi q\right)}{e^{-2 \pi q}\left(\pi/2\right)^2}$
can of course be absorbed via a suitable redefinition of $\hat\phi(q)$.

\subsection{Quantization and matrix version}

The above discussion regarding the eigenmodes of  the d'Alembertian  \eqref{Box-expression} is completely classical, even though we are claiming to work in the framework of matrix models. This may seem suspect, but it can be fully justified as follows. Since the underlying fuzzy hyperboloid $H^4_n$ is a quantized coadjoint orbit, there is an $SO(4,1)$-equivariant quantization map\footnote{Strictly speaking, the underlying space is twistor space $\C P^{1,2}$, and we are restricting ourselves to the lowest sector of scalar modes here \cite{Sperling:2019xar}.} 
\begin{align}
\cQ:\quad \cC(H^4) \to \End(\cH)
\end{align}
which establishes an isometric equivalence between  commutative and the fuzzy "functions", and maps the matrix d'Alembertian $\Box$ to the above semi-classical version. Therefore all the eigenmodes and eigenvalues computed in the classical case carry over without corrections to the fuzzy case, {\em in the free theory}. This justifies the classical computations in this work.

\section{Path integral quantization}
\label{sec:path-integral}

In this section we compute the propagator of a scalar field $\phi(x)$ in the FLRW geometry \eqref{eff-metric-FRW}. After having investigated the action   in Sec. \ref{Sec:action}, the propagator of $\phi(x)$ is explicitly worked out in Sec. \ref{Sec:propagator-mom-pos-space}.

\subsection{The action} \label{Sec:action}

In the semi-classical limit, the action for a scalar field $\phi(x)$ having mass $m$ can be written as
\begin{align}
S_{\varepsilon}\left[\phi\right]= \int \Omega   \phi^*(x)\left(-\Box -m^2 + i \varepsilon \right)\phi(x), \label{action-1}
\end{align}
where the expression of $\Omega$ can read from Eq. \eqref{symplectic-volume-form} and, as usual, $\varepsilon$ is a small positive number which should be let tend to zero after integration. The exact matrix version of the action has the same form with the trace $\Tr$ replacing the integral $\int d\Omega$, and the $i\varepsilon$ term  ensures that the matrix path integral $\int D\phi \, e^{i S}$ is well-defined \cite{Steinacker:2019fcb}.
 We can evaluate $S_{\varepsilon}\left[\phi\right]$ by employing the following decomposition of $\phi(x)$ in the basis of eigenmodes \eqref{Upsilon}:
\begin{align}
\phi(x)=  \sum_{l,m}\int ds dq \Bigl[\phi^+_{s,q,l,m} \Upsilon^{s_{+},q}_{l,m}(x)+\phi^-_{s,q,l,m} \Upsilon^{s_{-},q}_{l,m}(x)  \Bigr],
\label{phi-decomposition}
\end{align}
$\phi^+_{s,q,l,m},\phi^-_{s,q,l,m}$ being the coefficients of such a decomposition. In order to ease the notation, in our forthcoming calculations we will set
\begin{align}
\phi^{\pm} &\equiv \phi^{\pm}_{s,q,l,m}\ , 
\nonumber \\
\phi^{\prime \pm} &\equiv \phi^{\pm}_{s^\prime,q^\prime,l^\prime,m^\prime}\ . 
\label{compact-notation-phi-plus-minus-1}
\end{align}
Bearing in mind Eqs. \eqref{lambda-expression} and \eqref{phi-decomposition}, the action \eqref{action-1} becomes
\begin{align}
 S_{\varepsilon}\left[\phi\right]&= \sum_{l,m}\int   ds dq \left[-\dfrac{1}{R^2}\left(q^2-s^2-\dfrac{5}{4}\right)-m^2 + i \varepsilon \right]
 \nonumber \\
& \times \int \Omega  \phi^*(x) \left[\left(\phi^+\right) \Upsilon^{s_{+},q}_{l,m}(x)+\left(\phi^-\right) \Upsilon^{s_{-},q}_{l,m}(x) \right]
\nonumber \\
&= \sum_{l,m}\sum_{l^\prime,m^\prime}\int   ds dq ds^\prime dq^\prime \left[-\dfrac{1}{R^2}\left(q^2-s^2-\dfrac{5}{4}\right)-m^2 + i \varepsilon \right] 
\nonumber \\
& \times \int \Omega  \left[\left(\phi^{\prime +}\right) \Upsilon^{s^\prime_{+},q^\prime}_{l^\prime,m^\prime}(x)+\left(\phi^{\prime -}\right) \Upsilon^{s^\prime_{-},q^\prime}_{l^\prime,m^\prime}(x) \right]^* \left[\left(\phi^+\right) \Upsilon^{s_{+},q}_{l,m}(x)+\left(\phi^- \right) \Upsilon^{s_{-},q}_{l,m}(x) \right].
\end{align}
Thanks to the orthogonality relations \eqref{orthogonality-relations-final}, the above equation gives
\begin{align}
   S_{\varepsilon}\left[\phi\right]&=  \sum_{l,m}\sum_{l^\prime,m^\prime}\int   ds dq ds^\prime dq^\prime \left[-\dfrac{1}{R^2}\left(q^2-s^2-\dfrac{5}{4}\right)-m^2 + i \varepsilon \right] \Biggl[  \left(\phi^{\prime +}\right)^* \phi^+ \langle \Upsilon^{s^\prime_{+},q^\prime}_{l^\prime,m^\prime},  \Upsilon^{s_{+},q}_{l,m} \rangle 
\nonumber \\   
&+ \left(\phi^{\prime +}\right)^* \phi^- \langle \Upsilon^{s^\prime_{+},q^\prime}_{l^\prime,m^\prime},  \Upsilon^{s_{-},q}_{l,m} \rangle + \left(\phi^{\prime -}\right)^* \phi^+ \langle \Upsilon^{s^\prime_{-},q^\prime}_{l^\prime,m^\prime},  \Upsilon^{s_{+},q}_{l,m} \rangle +\left(\phi^{\prime -}\right)^* \phi^- \langle \Upsilon^{s^\prime_{-},q^\prime}_{l^\prime,m^\prime},  \Upsilon^{s_{-},q}_{l,m} \rangle   \Biggr]
\nonumber \\
&= \sum_{l,m}\sum_{l^\prime,m^\prime}\int   ds dq ds^\prime dq^\prime \Biggl\{ \left[-\dfrac{1}{R^2}\left(q^2-s^2-\dfrac{5}{4}\right)-m^2 + i \varepsilon \right] \dfrac{e^{-2\pi q}(\pi/2)^2 \, \delta_{l l^\prime} \delta_{m m^\prime}}{q \sinh(\pi q)} \delta(q-q^\prime) 
\nonumber \\
& \times \delta(s-s^\prime) \Bigl[\left(\phi^{\prime +}\right)^* \left(\phi^+\right) b(q,s)+\left(\phi^{\prime +}\right)^* \left(\phi^-\right) a(q,-s)+\left(\phi^{\prime -}\right)^* \left(\phi^+\right) a(q,s)
\nonumber \\
&+\left(\phi^{\prime -}\right)^* \left(\phi^-\right) b(q,-s)\Bigr]\Biggr\},
\end{align}
where in the last passage we have exploited the fact that  all terms proportional to $\delta(s+s^\prime)$ give a vanishing contribution, since both $s$ and $s^\prime$ are positive (cf. Eq. \eqref{s-equals-vert-mu}). If  we introduce the square matrix
\begin{align}
    \mathscr{B}(q,s)=
    \begin{bmatrix}
     b(q,s) & a(q,-s)\\
     a(q,s) & b(q,-s)
    \end{bmatrix},
\end{align}
then we can write 
\begin{align}
    \begin{bmatrix}
 \left(\phi^{\prime +}\right)^* & \left(\phi^{\prime -}\right)^*
\end{bmatrix}
\mathscr{B}(q,s)  \begin{bmatrix}
 \phi^+ \\ 
 \phi^-
\end{bmatrix}
&=\left(\phi^{\prime +}\right)^* \left(\phi^+\right) b(q,s)+\left(\phi^{\prime +}\right)^* \left(\phi^-\right) a(q,-s)
\nonumber \\
&+\left(\phi^{\prime -}\right)^* \left(\phi^+\right) a(q,s)
+\left(\phi^{\prime -}\right)^* \left(\phi^-\right) b(q,-s),
\end{align}
and hence, in conclusion, the action assumes the form
\begin{align}
     S_{\varepsilon}\left[\phi\right] &=   \sum_{l,m}\sum_{l^\prime,m^\prime}\int   ds dq ds^\prime dq^\prime  \left[-\dfrac{1}{R^2}\left(q^2-s^2-\dfrac{5}{4}\right)-m^2 + i \varepsilon \right] \dfrac{e^{-2\pi q}(\pi/2)^2 }{q \sinh(\pi q)} 
\nonumber \\
&\times \delta_{l l^\prime} \delta_{m m^\prime} \delta(q-q^\prime) \delta(s-s^\prime)  \begin{bmatrix}
 \left(\phi^{\prime +}\right)^* & \left(\phi^{\prime -}\right)^*
\end{bmatrix}
\mathscr{B}(q,s)  \begin{bmatrix}
 \phi^+ \\ 
 \phi^-
\end{bmatrix}.
\label{action-final-expression}
\end{align}

\subsection{The propagator in momentum space and position space} \label{Sec:propagator-mom-pos-space}

In the following, we  evaluate the propagator of the massive scalar field $\phi(x)$. We first provide the 
general expressions for  the propagator both in momentum space and position space. 
The late-time propagator is then elaborated in Sec. \ref{Sec:prop-eta-inf}, recalling also a related topic -- i.e. the expansion of plane waves in terms of spherical harmonics -- in Sec. \ref{sec:Plane-wave-expansion}. Finally, in Sec. \ref{Sec:prop-early-time} we compute the propagator across the BB. 

Starting from the action \eqref{action-final-expression} and adopting the compact notation (cf. Eq. \eqref{compact-notation-phi-plus-minus-1})
\begin{align}
\Phi^{\pm} &\equiv \begin{bmatrix}
 \phi^+ \\ 
 \phi^-
\end{bmatrix},
\nonumber \\
  \left( \Phi^{\prime \pm}\right)^{\dagger} &\equiv \begin{bmatrix}
 \left(\phi^{\prime +}\right)^* & \left(\phi^{\prime -}\right)^*
\end{bmatrix},
\end{align}
the propagator in momentum space reads as
\begin{align}
\left \langle \left(\Phi^{\pm}\right)   \left( \Phi^{\prime \pm}\right)^{\dagger} \right \rangle &= \delta_{l l^\prime} \delta_{m m^\prime} \delta\left(q-q^\prime\right)
\delta\left(s-s^\prime\right) \dfrac{1}{\dfrac{1}{R^2}\left(s^2 - q^2 +\dfrac{5}{4}\right)-m^2 + i \varepsilon} 
\nonumber \\
& \times \dfrac{4q \sinh (\pi q)}{e^{-2 \pi q}\pi^2} \left[\mathscr{B}(q,s)\right]^{-1},
\label{propagator-momentum-space}
\end{align}
where
\begin{align}
\left[\mathscr{B}(q,s)\right]^{-1} =\dfrac{1}{\det \left[\mathscr{B}(q,s)\right]} \begin{bmatrix}
b(q,-s) & -a(q,-s) \\
-a(q,s) & b(q,s)
\end{bmatrix},
\end{align}
and (see Eq. \eqref{a-q-s-and-b-q-s})
\begin{align}
    \det \left[\mathscr{B}(q,s)\right]= \dfrac{2}{s^2} \left[\cosh\left(2 \pi q\right)+\cosh\left(2 \pi s\right)\right].
\end{align}

Therefore, bearing in mind Eq. \eqref{propagator-momentum-space}, the local propagator in position space reads as
\begin{align}
  \langle \phi(x) \phi^*(x^\prime)\rangle &= \sum_{l,m}\sum_{l^\prime,m^\prime} \int ds dq ds^\prime dq^\prime 
\begin{bmatrix}
 \Upsilon^{s_+,q}_{l,m} (x) & \Upsilon^{s_-,q}_{l,m}(x)
\end{bmatrix}  
  \left \langle \left(\Phi^{\pm}\right)   \left( \Phi^{\prime \pm}\right)^{\dagger} \right \rangle
  \begin{bmatrix}
  \left( \Upsilon^{s^\prime_{+},q^\prime}_{l^\prime,m^\prime} \left(x^\prime\right)\right)^* \\
   \left( \Upsilon^{s^\prime_{-},q^\prime}_{l^\prime,m^\prime} \left(x^\prime\right)\right)^* 
  \end{bmatrix}.
\label{propagator-position-space}  
\end{align}

\subsubsection{The propagator in the flat regime and with $\eta \to + \infty$}\label{Sec:prop-eta-inf}

We are  mainly interested in the local propagator for distances far below the curvature scale,  but
keeping the oscillating nature of the modes in the late-time regime $\eta \to +\infty$. This is the flat regime FR defined in \eq{regime-A}, where the eigenmodes \eqref{Upsilon}  reduce to \eqref{flate-regime-Upsilon}.
It then follows from  Eq. \eqref{propagator-position-space}  that the  late-time  local propagator can be written as the sum of a leading piece and a subleading part, i.e., 
\begin{align}
\langle \phi(x) \phi^*(x^\prime)\rangle \overunderset{\eta \to + \infty}{{\rm FR}}{\sim} \,  \langle \phi(x) \phi^*(x^\prime)\rangle^{\eta \to + \infty, {\rm FR}}_{\rm L} +\langle \phi(x) \phi^*(x^\prime)\rangle^{\eta \to + \infty, {\rm FR}}_{\rm SL},
\label{flat-regime-propagator-leading&subleading}     
\end{align}
where \qm{L} and \qm{SL} stand for \qm{leading} and \qm{subleading}, respectively. 
The leading term reads as\footnote{Notice that if we would keep here only the asymptotic behavior of the spherical Bessel functions $j_l$ rather than their full form, the propagator would be ill-defined, as the sum over $l,m$ would lead to a singular dependence on $\theta,\varphi$.}
\begin{align}
\langle \phi(x) \phi^*(x^\prime)\rangle^{\eta \to + \infty,{\rm FR}}_{\rm L} &= \dfrac{2R^2}{\pi^3}  \sum_{l,m} \dfrac{ Y^m_l(\theta,\varphi)\left[Y^m_l(\theta^\prime,\varphi^\prime)\right]^*}{\sqrt{\left(\cosh^3 \eta\right)\left(\cosh^3 \eta^\prime\right)}} \int ds dq
\dfrac{j_l\left(q \chi\right)j_l\left(q \chi^\prime\right)}{q^{2l}\left(s^2 - q^2 + \dfrac{5}{4}-m^2 R^2 + i \varepsilon \right)} 
\nonumber \\
& \times 2 \cos \left[s\left(\eta - \eta^\prime\right)\right] \left \vert \Gamma\left(iq+l+1\right) \right \vert^2 q \sinh(\pi q).
\label{propagator-flat-regime-leading-1}
\end{align}
The above equation can be simplified by means of the identity
\begin{align}
\left \vert \Gamma(iq+l+1) \right \vert^2 = \dfrac{\pi }{ q \sinh (\pi q)} \prod_{n=0}^{l} \left[q^2 + \left(l-n\right)^2\right],
\label{modulus-Gamma-formula}
\end{align}
which can be proved via the recurrence relation $\Gamma(1+z)=z \Gamma(z)$ \cite{Abramowitz-Stegun(1964)} along with Eq. \eqref{Gamma-identity-1}. Moreover,  the formula \eqref{modulus-Gamma-formula} leads to (cf. \eqref{regime-A})
\begin{align}
 q \sinh(\pi q)  \left \vert \Gamma\left(iq+l+1\right) \right \vert^2  \overset{q \gg l}{\sim } \pi \,  q^{2l+2}.
 \label{modulus-Gamma-formula-2}
\end{align}
The restriction to $q \gg l$ means that we ignore the extreme IR regime of the propagator, which is justified for the typical applications of (quantum) field theory.

Therefore, by exploiting \eqref{modulus-Gamma-formula-2} and after some calculation, Eq. \eqref{propagator-flat-regime-leading-1} becomes
\begin{align}
\langle \phi(x) \phi^*(x^\prime)\rangle^{\eta \to + \infty,{\rm FR}}_{\rm L} &= \dfrac{2R^2}{\pi^2}  \sum_{l,m} \dfrac{ Y^m_l(\theta,\varphi)\left[Y^m_l(\theta^\prime,\varphi^\prime)\right]^*}{\sqrt{\left(\cosh^3 \eta\right)\left(\cosh^3 \eta^\prime\right)}} \int^{+ \infty}_{-\infty} ds\, e^{i s\left(\eta - \eta^\prime\right)}
\nonumber \\
& \times \int^{+ \infty}_{-\infty} dq
\dfrac{q^2 j_l\left(q \chi\right)j_l\left(q \chi^\prime\right)}{\left(s^2 - q^2 + \dfrac{5}{4} -m^2 R^2 + i \varepsilon \right)}  
\nn \\
&=\dfrac{4R^2}{\pi^2}  \sum_{l,m} \dfrac{ Y^m_l(\theta,\varphi)\left[Y^m_l(\theta^\prime,\varphi^\prime)\right]^*}{\sqrt{\left(\cosh^3 \eta\right)\left(\cosh^3 \eta^\prime\right)}} \int^{+ \infty}_{-\infty} ds\, e^{i s\left(\eta - \eta^\prime\right)}
\nonumber \\
& \times \int^{+ \infty}_{0} dq
\dfrac{q^2 j_l\left(q \chi\right)j_l\left(q \chi^\prime\right)}{\left(s^2 - q^2 + \dfrac{5}{4}-m^2 R^2 + i \varepsilon \right)}, 
\label{propagator-flat-regime-leading-2}
\end{align}
where we have taken into account that the integrand is an even function of $q$ due to the property $j_l(-x)=(-1)^l j_l(x)$. 
Up to the $\eta$-dependent normalization factor which reflects the cosmic expansion, we recover precisely the Feynman propagator on a flat 3+1-dimensional Minkowski space, including the appropriate $i\varepsilon$ prescription which ensures local causality and determines the arrow of time along growing $\eta$.

The subleading contribution occurring in Eq. \eqref{flat-regime-propagator-leading&subleading} is
\begin{align}
\langle \phi(x) \phi^*(x^\prime)\rangle^{\eta \to + \infty,{\rm FR}}_{\rm SL} &= \dfrac{4R^2}{\pi^5}  \sum_{l,m} \dfrac{ Y^m_l(\theta,\varphi)\left[Y^m_l(\theta^\prime,\varphi^\prime)\right]^*}{\sqrt{\left(\cosh^3 \eta\right)\left(\cosh^3 \eta^\prime\right)}} \int ds dq
\dfrac{j_l\left(q \chi\right)j_l\left(q \chi^\prime\right)s  \sinh(\pi s)}{q^{2l}\left(s^2 - q^2 + \dfrac{5}{4} -m^2 R^2 + i \varepsilon \right)} 
\nonumber \\
&\times q \sinh(\pi q) \cosh(\pi q) \left \vert \Gamma\left(iq+l+1\right) \right \vert^2   
\nn \\
& \times {\rm Re} \left[e^{is\left(\eta+\eta^\prime\right)}\Gamma\left(\dfrac{1}{2}-iq-is\right)\Gamma\left(\dfrac{1}{2}+iq-is\right)\Gamma^2\left(is\right)\right].
\end{align}
By exploiting Eq.  \eqref{modulus-Gamma-formula-2} and after some arrangement, we find
\begin{align}
\langle \phi(x) \phi^*(x^\prime)\rangle^{\eta \to + \infty,{\rm FR}}_{\rm SL} &= \dfrac{2R^2}{\pi^4}  \sum_{l,m} \dfrac{ Y^m_l(\theta,\varphi)\left[Y^m_l(\theta^\prime,\varphi^\prime)\right]^*}{\sqrt{\left(\cosh^3 \eta\right)\left(\cosh^3 \eta^\prime\right)}} \int^{+ \infty}_{-\infty} ds \, e^{is\left(\eta+\eta^\prime\right)} 
\nn \\
& \times \int^{+ \infty}_{-\infty}dq
\dfrac{j_l\left(q \chi\right)j_l\left(q \chi^\prime\right)q^2 s  \cosh(\pi q)   \sinh(\pi s)}{\left(s^2 - q^2 + \dfrac{5}{4} -m^2 R^2 + i \varepsilon \right)} 
\nonumber \\
& \times  \Gamma\left(\dfrac{1}{2}-iq-is\right)\Gamma\left(\dfrac{1}{2}+iq-is\right)\Gamma^2\left(is\right)
\nn \\
&=\dfrac{4R^2}{\pi^4}  \sum_{l,m} \dfrac{ Y^m_l(\theta,\varphi)\left[Y^m_l(\theta^\prime,\varphi^\prime)\right]^*}{\sqrt{\left(\cosh^3 \eta\right)\left(\cosh^3 \eta^\prime\right)}} \int^{+ \infty}_{-\infty} ds \, e^{is\left(\eta+\eta^\prime\right)} 
\nn \\
& \times \int^{+ \infty}_{0}dq
\dfrac{j_l\left(q \chi\right)j_l\left(q \chi^\prime\right)q^2 s  \cosh(\pi q)   \sinh(\pi s)}{\left(s^2 - q^2 + \dfrac{5}{4} -m^2 R^2 + i \varepsilon \right)} 
\nonumber \\
& \times  \Gamma\left(\dfrac{1}{2}-iq-is\right)\Gamma\left(\dfrac{1}{2}+iq-is\right)\Gamma^2\left(is\right).
\label{propagator-flat-regime-subleading}
\end{align}

The last expression depends on the sum $\eta + \eta^\prime$ and hence it is rapidly oscillating  at late times (i.e., for $\eta, \eta^\prime \to +\infty$). This justifies the fact that Eq.  \eqref{propagator-flat-regime-subleading} gives a subleading contribution with respect to Eq. \eqref{propagator-flat-regime-leading-2}, which, on the contrary, depends on the difference $\eta - \eta^\prime$. However, a crucial consistency check consists in proving that the leading term \eqref{propagator-flat-regime-leading-2} has, apart from the normalization factor, the form of the usual local Feynman propagator on a flat four-dimensional spacetime. This task is fulfilled in the  Sec. \ref{sec:Plane-wave-expansion}.  

\subsubsection{Plane-wave expansion in terms of the spherical harmonics}\label{sec:Plane-wave-expansion}

The scalar plane waves can be expanded in $\R^3$ in terms of spherical harmonics via the   Rayleigh equation \cite{Newton1982}
\begin{align}
    e^{i \vec q \cdot \vec x} = 
    4\pi\sum\limits_{l,m} i^l j_l(q r) Y_{l}^m(\hat q) \left[Y_{l}^m(\hat x)\right]^*,
\end{align}
where $q=|\vec q|, \ r=|\vec x|$,
$\hat x = \frac {\vec x}r$, $\hat q = \frac {\vec q}q$, and $j_l(q r)$ are the spherical Bessel functions \cite{Abramowitz-Stegun(1964)}. In the above equation,  the complex conjugation can be interchanged between the two spherical harmonics due to the symmetry of the scalar product $ \vec q \cdot \vec x$. Therefore, we have
\begin{align}
e^{i \vec q \cdot (\vec x - \vec x^\prime)} 
&=  (4\pi)^2\sum\limits_{l,m} \sum\limits_{l^\prime,m^\prime} \left(i\right)^{l-l^\prime} j_l(q r) j_{l^\prime}\left(q r^\prime\right) Y_{l}^m(\hat q)\left[Y_{l^\prime}^{m^\prime}(\hat q)\right]^* \left[Y_{l}^m(\hat x)\right]^* Y_{l^\prime}^{m^\prime}(\hat x^\prime),
\end{align}
recalling that $Y_{l}^m(-\hat x^\prime) = (-1)^l Y_{l}^m(\hat x^\prime)$. 
Now we integrate this relation over all $\vec q$ with fixed radius $q$:
\begin{align}
\int\limits_{S_q^2}d\Omega_q
e^{i \vec q \cdot (\vec x - \vec x^\prime)} 
&=  (4\pi)^2\sum\limits_{l,m} \sum\limits_{l^\prime,m^\prime} \left(i\right)^{l-l^\prime} j_l(q r) j_{l^\prime}(q r^\prime) \left[Y_{l}^m(\hat x)\right]^*Y_{l^\prime}^{m^\prime}(\hat x^\prime)   \int\limits_{S_q^2}d\Omega_q Y_{l}^m(\hat q)\left[Y_{l^\prime}^{m^\prime}(\hat q)\right]^*
\nonumber \\
&=  (4\pi)^2\sum\limits_{l,m}  j_l(q r) j_{l}(q r^\prime)
\left[Y_{l}^m(\hat x)\right]^*  Y_{l}^{m}(\hat x^\prime),
\end{align}
using the orthogonality of the spherical harmonics. Therefore, we can write e.g. the propagator in the form  
\begin{align}
\int d^3q\,  \frac{e^{i \vec q \cdot (\vec x - \vec x^\prime)}}{s^2 - q^2-M^2} &=  \int_0^{+\infty} dq \ q^2 \int\limits_{S_q^2}d\Omega_q \frac{e^{i \vec q \cdot (\vec x - \vec x^\prime)}}{s^2 - q^2 -M^2} 
\nonumber \\
&= (4\pi)^2 \sum\limits_{l,m} \left[Y_{l}^m(\hat x)\right]^*  Y_{l}^{m}(\hat x^\prime) \int_0^{+\infty} dq  \frac{q^2}{s^2 - q^2-M^2} j_l(q r) j_{l}(q r^\prime).
\label{integral-bessel-propagator}     
\end{align}
The analogy with Eq. \eqref{propagator-flat-regime-leading-2} should be noted. 

\subsubsection{The propagator in the flat regime and with $\eta \to 0$} \label{Sec:prop-early-time}

The behaviour  of the scalar field (cf. Eq. \eqref{action-1}) near the BB  can be inferred from the form assumed by its propagator \eqref{propagator-position-space} for small times, i.e., when $\eta \to 0$. Here, we will also employ the flat regime \eqref{regime-A}.

The definition of the associated Legendre function of the first kind $ \mathsf{P}^{\mu}_{\nu}\left(x\right)$ having $x \in (-1,1)$  \cite{Abramowitz-Stegun(1964)}
\begin{align}
    \mathsf{P}^{\mu}_{\nu}\left(x\right)= \dfrac{1}{\Gamma(1-\mu)}\left(\frac{1+x}{1-x}\right)^{\mu/2}
{}_{2}\mathrm{F}_{1}\left(-\nu,\nu+1;1-\mu;\tfrac{1}{2}-\tfrac{1}{2}x\right),  
\end{align}
jointly with Eq. \eqref{degree-nu-with-plus}, leads to the expression
\begin{align}
\mathsf{P}^{\pm is}_{\nu}\left(\tanh \eta\right) &=  \dfrac{{}_{2}\mathrm{F}_{1}\left(\dfrac{1}{2}-i  q ,\dfrac{1}{2}+i  q ; 1 \mp is;\dfrac{1}{2}-\dfrac{1}{2} \tanh \eta \right)}{\Gamma(1 \mp is)} \left(\dfrac{1+ \tanh \eta}{1- \tanh \eta}\right)^{\pm is/2},
\label{P-is-nu-tanh-eta}
\end{align}
where we have exploited the  relation ${}_{2}\mathrm{F}_{1}(a,b;c;x)={}_{2}\mathrm{F}_{1}(b,a;c;x)$ to  write
\begin{align}
   {}_{2}\mathrm{F}_{1}\left(\dfrac{1}{2}-i  \vert q\vert  ,\dfrac{1}{2}+i  \vert q\vert  ; 1 \mp is;\dfrac{1}{2}-\dfrac{1}{2} \tanh \eta \right) ={}_{2}\mathrm{F}_{1}\left(\dfrac{1}{2}-i  q ,\dfrac{1}{2}+i  q ; 1 \mp is;\dfrac{1}{2}-\dfrac{1}{2} \tanh \eta \right). 
\end{align}
The  expansion about $\eta=0$ of the hypergeometric function occurring above  can be written as
\begin{align}
{}_{2}\mathrm{F}_{1}\left(\dfrac{1}{2}-i  q ,\dfrac{1}{2}+i  q ; 1 \mp is;\dfrac{1}{2}-\dfrac{1}{2} \tanh \eta \right) &=  {}_{2}\mathrm{F}_{1}\left(\dfrac{1}{2}-i  q ,\dfrac{1}{2}+i  q ; 1 \mp is;\dfrac{1}{2}\right) + {\rm O}\left(\eta\right)
\nn \\
&= \dfrac{\sqrt{\pi} \left(2\right)^{\pm is} \Gamma\left( 1 \mp is \right)}{\Gamma \left[\dfrac{1}{2}\left(\dfrac{3}{2}-iq \mp is\right)\right]\Gamma \left[\dfrac{1}{2}\left(\dfrac{3}{2}+iq \mp is\right)\right]}
\nn \\
&+{\rm O}\left(\eta\right),
\label{expansion-eta-zero-hypergeometric}
\end{align}
where we have exploited Eq. (15.1.26) in Ref. \cite{Abramowitz-Stegun(1964)}. Therefore, from Eqs. \eqref{P-is-nu-tanh-eta} and \eqref{expansion-eta-zero-hypergeometric}, we find 
\begin{align}
\mathsf{P}^{\pm is}_{\nu}\left(\tanh \eta\right) &\overset{\eta \to 0}{\sim}  \,
e^{\pm i\eta s} \,\dfrac{\sqrt{\pi} \left(2\right)^{\pm is}}{\Gamma \left[\dfrac{1}{2}\left(\dfrac{3}{2}-iq \mp is\right)\right]\Gamma \left[\dfrac{1}{2}\left(\dfrac{3}{2}+iq \mp is\right)\right]}.
\end{align}

Thus, from the last relation jointly with Eq. \eqref{Legendre-bessel-flat-complete}, the early-time  expression of the eigenfunctions \eqref{Upsilon} evaluated in the flat regime \eqref{regime-A} become
\begin{align}
\Upsilon^{s_{\pm},q}_{l,m}(\eta,\chi,\theta,\varphi)&\overunderset{\eta \to 0}{{\rm FR}}{\sim}\,\dfrac{\sqrt{\pi}\,2^{\pm is}\, e^{\pm i \eta s} }{\Gamma \left(\dfrac{3}{4}-\dfrac{iq}{2} \mp \dfrac{is}{2}\right)\Gamma \left(\dfrac{3}{4}+\dfrac{iq}{2} \mp \dfrac{is}{2}\right)}\dfrac{e^{-\pi q} j_l\left(q \chi\right) \Gamma(iq+l+1)}{ q^l} Y^m_l \left(\theta,\varphi\right).
\label{upsilon-flat-regime-small-time}
\end{align}
As a consequence of Eqs. \eqref{propagator-position-space} and \eqref{upsilon-flat-regime-small-time}, we find that the early-time propagator can be written as 
\begin{align}
\langle \phi(x) \phi^*(x^\prime)\rangle \overunderset{\eta \to 0}{{\rm FR}}{\sim} \langle \phi(x) \phi^*(x^\prime)\rangle^{\rm FR}_{\rm ET1} + \langle \phi(x) \phi^*(x^\prime)\rangle^{\rm FR}_{\rm ET2},
\label{propagator-FR-small-eta-1}
\end{align}
where \qm{ET} means \qm{early-time} propagator. The first term  reads as
\begin{align}
\langle \phi(x) \phi^*(x^\prime)\rangle^{\rm FR}_{\rm ET1}&=- 8  R^2 \sum_{l,m} Y^m_l(\theta,\varphi)\left[Y^m_l(\theta^\prime,\varphi^\prime)\right]^* \int ds dq\, \dfrac{j_l \left( q \chi\right)j_l \left( q \chi^\prime \right)}{q^{2l}} 
\nn \\
& \times    \dfrac{ s \,q\, \cosh(\pi q)  \sinh (\pi q) \left \vert \Gamma(iq+l+1) \right \vert^2}{\sinh(\pi s) \left(s^2 - q^2 + \dfrac{5}{4}-m^2 R^2 + i \varepsilon \right)\left[\cosh(2 \pi q)+\cosh(2 \pi s)\right]} 
\nn \\
& \times {\rm Re} \left[ \dfrac{2^{-2is} \; e^{-is \left(\eta + \eta^\prime \right)}\;\Gamma^{-1}\left(\dfrac{1}{2}-iq-is\right)\Gamma^{-1}\left(\dfrac{1}{2}+iq-is\right)}{\Gamma^2\left(\dfrac{3}{4}-\dfrac{iq}{2}+\dfrac{is}{2}\right)\Gamma^2\left(\dfrac{3}{4}+\dfrac{iq}{2}+\dfrac{is}{2}\right)} \right].
\end{align}
After some calculations, the leading contribution becomes
\begin{align}
\langle \phi(x) \phi^*(x^\prime)\rangle^{\rm FR}_{\rm ET1}&=- 4  R^2 \sum_{l,m} Y^m_l(\theta,\varphi)\left[Y^m_l(\theta^\prime,\varphi^\prime)\right]^* \int^{+ \infty}_{- \infty} ds\; 2^{2is} \; e^{is \left(\eta + \eta^\prime \right)} 
\nn \\
& \times \int^{+ \infty}_{- \infty} dq\, \dfrac{j_l \left( q \chi\right)j_l \left( q \chi^\prime \right)}{q^{2l}}    \dfrac{ q   \sinh (\pi q) \left \vert \Gamma(iq+l+1) \right \vert^2}{ \left(s^2 - q^2 + \dfrac{5}{4} -m^2 R^2 + i \varepsilon \right)} 
\nn \\
& \times \dfrac{ s \cosh(\pi q) }{\sinh(\pi s) \left[\cosh(2 \pi q)+\cosh(2 \pi s)\right]}  \dfrac{ \Gamma^{-1}\left(\dfrac{1}{2}+iq+is\right)\Gamma^{-1}\left(\dfrac{1}{2}-iq+is\right)}{\Gamma^2\left(\dfrac{3}{4}+\dfrac{iq}{2}-\dfrac{is}{2}\right)\Gamma^2\left(\dfrac{3}{4}-\dfrac{iq}{2}-\dfrac{is}{2}\right)} 
\nn \\
&=- 8 \pi  R^2 \sum_{l,m} Y^m_l(\theta,\varphi)\left[Y^m_l(\theta^\prime,\varphi^\prime)\right]^* \int^{+ \infty}_{- \infty} ds\; 2^{2is} \; e^{is \left(\eta + \eta^\prime \right)} 
\nn \\
& \times \int^{+ \infty}_{0} dq\,    \dfrac{ q^2 j_l \left( q \chi\right)j_l \left( q \chi^\prime \right) }{ \left(s^2 - q^2 + \dfrac{5}{4} -m^2 R^2 + i \varepsilon \right)} \dfrac{ s \cosh(\pi q) }{\sinh(\pi s) \left[\cosh(2 \pi q)+\cosh(2 \pi s)\right]} 
\nn \\
& \times  \dfrac{ \Gamma^{-1}\left(\dfrac{1}{2}+iq+is\right)\Gamma^{-1}\left(\dfrac{1}{2}-iq+is\right)}{\Gamma^2\left(\dfrac{3}{4}+\dfrac{iq}{2}-\dfrac{is}{2}\right)\Gamma^2\left(\dfrac{3}{4}-\dfrac{iq}{2}-\dfrac{is}{2}\right)}, 
\label{propgator-early-time-factor1-1}
\end{align}
where we have exploited Eq. \eqref{modulus-Gamma-formula-2} and the fact that the integrand is an even function of $q$. If we exploit the following relations \cite{Abramowitz-Stegun(1964)}
\begin{align}
\Gamma(2z)&= \dfrac{2^{2z-1}}{\sqrt{\pi}} \Gamma(z) \Gamma(z+1/2),
\nn \\   
\Gamma\left(1/4+iy\right)\Gamma\left(3/4-i
y\right)&=\frac{\pi\sqrt{2}}{\cosh\left(\pi y\right)+i\sinh\left(\pi y 
\right)},
\end{align}
then we can simplify the last term occurring in Eq. \eqref{propgator-early-time-factor1-1} as
\begin{align}
\dfrac{ \Gamma^{-1}\left(\dfrac{1}{2}+iq+is\right)\Gamma^{-1}\left(\dfrac{1}{2}-iq+is\right)}{\Gamma^2\left(\dfrac{3}{4}+\dfrac{iq}{2}-\dfrac{is}{2}\right)\Gamma^2\left(\dfrac{3}{4}-\dfrac{iq}{2}-\dfrac{is}{2}\right)}  = \dfrac{\cosh(\pi q) + i \sinh(\pi s)}{\pi \left(2^{2is}\right) \left\vert \Gamma\left(\dfrac{3}{4}+\dfrac{iq}{2}+\dfrac{is}{2}\right)\Gamma\left(\dfrac{3}{4}+\dfrac{iq}{2}-\dfrac{is}{2}\right) \right\vert^2},  
\label{Gamma-squared-modulus-squared-relation}
\end{align}
and hence, after some algebra, we get
\begin{align}
\langle \phi(x) \phi^*(x^\prime)\rangle^{\rm FR}_{\rm ET1}&=- 4  R^2 \sum_{l,m} Y^m_l(\theta,\varphi)\left[Y^m_l(\theta^\prime,\varphi^\prime)\right]^* \int\limits^{+ \infty}_{- \infty} ds     \int\limits^{+ \infty}_{0} dq\,    \dfrac{ q^2 j_l \left( q \chi\right)j_l \left( q \chi^\prime \right) e^{is \left(\eta + \eta^\prime \right)}  }{ \left(s^2 - q^2 + \dfrac{5}{4} -m^2 R^2 + i \varepsilon \right)}
\nn \\
& \times \dfrac{ s \cosh(\pi q) }{\sinh(\pi s) \left[\cosh( \pi q)-i\sinh( \pi s)\right]} \dfrac{1}{\left\vert \Gamma\left(\dfrac{3}{4}+\dfrac{iq}{2}+\dfrac{is}{2}\right)\Gamma\left(\dfrac{3}{4}+\dfrac{iq}{2}-\dfrac{is}{2}\right) \right\vert^2 }.
\label{early-time-term1}
\end{align}
We can obtain an estimate for the last term appearing in the above equation. Indeed, taking into account the relation \cite{Nist}
\begin{align}
    \left \vert  \Gamma(x+iy) \right \vert^2 \geq \dfrac{\Gamma^2(x)}{\cosh(\pi y)},
\end{align}
we obtain the inequality
\begin{align}
\dfrac{1}{\left\vert \Gamma\left(\dfrac{3}{4}+\dfrac{iq}{2}+\dfrac{is}{2}\right)\Gamma\left(\dfrac{3}{4}+\dfrac{iq}{2}-\dfrac{is}{2}\right) \right\vert^2 } \leq \dfrac{1}{2\, \Gamma^{4}\left(\dfrac{3}{4}\right)} \left[\cosh(\pi q)+ \cosh(\pi s)\right],
\label{Gamma-estimate}
\end{align}
where $\Gamma^4\left(\tfrac{3}{4}\right) \approx 2.25$.

The second term appearing in Eq. \eqref{propagator-FR-small-eta-1} is
\begin{align}
\langle \phi(x) \phi^*(x^\prime)\rangle^{\rm FR}_{\rm ET2}&=\dfrac{8R^2}{\pi}\sum_{l,m} Y^m_l(\theta,\varphi)\left[Y^m_l(\theta^\prime,\varphi^\prime)\right]^*
\nn \\
& \times \int ds dq \, \dfrac{j_l \left(q \chi\right)j_l \left(q \chi^\prime\right)}{q^{2l}} \dfrac{\cos\left[s \left(\eta-\eta^\prime\right)\right]\, q \sinh(\pi q) \left \vert \Gamma(iq+l+1)\right \vert^2}{ \left(s^2 - q^2 + \dfrac{5}{4}-m^2 R^2 + i \varepsilon \right)}
\nn \\
& \times \dfrac{s \left[\sinh^2( \pi s)+\cosh^2( \pi q)\right]}{\sinh(\pi s)\left[\cosh(2 \pi q)+\cosh(2 \pi s)\right]}  \dfrac{1}{\left\vert \Gamma\left(\dfrac{3}{4}+\dfrac{iq}{2}+\dfrac{is}{2}\right)\Gamma\left(\dfrac{3}{4}+\dfrac{iq}{2}-\dfrac{is}{2}\right) \right\vert^2 },
\end{align}
and, upon taking into account Eq. \eqref{modulus-Gamma-formula-2} jointly with the fact that the integrand is an even function of  $q$, we obtain, after some algebra, 
\begin{align}
\langle \phi(x) \phi^*(x^\prime)\rangle^{\rm FR}_{\rm ET2}&=4R^2\sum_{l,m} Y^m_l(\theta,\varphi)\left[Y^m_l(\theta^\prime,\varphi^\prime)\right]^* \int\limits^{+ \infty}_{- \infty} ds     \int\limits^{+ \infty}_{0} dq\,    \dfrac{ q^2 j_l \left( q \chi\right)j_l \left( q \chi^\prime \right) e^{is \left(\eta - \eta^\prime \right)}  }{ \left(s^2 - q^2 + \dfrac{5}{4} -m^2 R^2 + i \varepsilon \right)}
\nn \\
& \times \dfrac{s}{\sinh(\pi s)} \dfrac{1}{\left\vert \Gamma\left(\dfrac{3}{4}+\dfrac{iq}{2}+\dfrac{is}{2}\right)\Gamma\left(\dfrac{3}{4}+\dfrac{iq}{2}-\dfrac{is}{2}\right) \right\vert^2 }.
\label{early-time-term2}
\end{align}
Therefore, if we sum the contributions  \eqref{early-time-term1} and \eqref{early-time-term2}, the early-time propagator \eqref{propagator-FR-small-eta-1} becomes
\begin{align}
\langle \phi(x) \phi^*(x^\prime)\rangle \overunderset{\eta \to 0}{{\rm FR}}{\sim}&\, 4R^2\sum_{l,m} Y^m_l(\theta,\varphi)\left[Y^m_l(\theta^\prime,\varphi^\prime)\right]^* \int\limits^{+ \infty}_{- \infty} ds     \int\limits^{+ \infty}_{0} dq\,    \dfrac{ q^2 j_l \left( q \chi\right)j_l \left( q \chi^\prime \right)   }{ \left(s^2 - q^2 + \dfrac{5}{4}-m^2 R^2 + i \varepsilon \right)}
\nn \\
& \times  \dfrac{1}{\left\vert \Gamma\left(\dfrac{3}{4}+\dfrac{iq}{2}+\dfrac{is}{2}\right)\Gamma\left(\dfrac{3}{4}+\dfrac{iq}{2}-\dfrac{is}{2}\right) \right\vert^2 } \frac{s}{\sinh(\pi s)}
\nn \\
& \times \left[e^{i s(\eta-\eta^\prime)} - 
\dfrac{\cosh(\pi q) e^{i s(\eta+\eta^\prime)}}{ \cosh( \pi q)-i\sinh( \pi s)}
\right].
\end{align}
The last term occurring in the above equation can be arranged as follows
\begin{align}
&\frac{s}{\sinh(\pi s)}
    \left[e^{i s(\eta-\eta^\prime)} - 
     \dfrac{\cosh(\pi q) e^{i s(\eta+\eta^\prime)}}{ \cosh( \pi q)-i\sinh( \pi s)}
     \right] 
\nonumber\\
&=\frac{s}{\sinh(\pi s)}
    \dfrac{\left[e^{i s(\eta-\eta^\prime)} - e^{i s(\eta+\eta^\prime)} \right]\cosh(\pi q) -i   \,e^{i s(\eta-\eta^\prime)} \sinh(\pi s)}{ \cosh( \pi q)-i\sinh( \pi s)}
\nonumber\\
 & \overset{\eta,\eta^\prime \to 0}{\sim} \,
    \dfrac{ -is }{\left[\cosh( \pi q)-i\sinh( \pi s)\right]} e^{i s (\eta - \eta')},
\end{align}
and hence, for  $\eta, \eta^\prime \to 0$, we get the final formula
\begin{align}
\langle \phi(x) \phi^*(x^\prime)\rangle\overunderset{\eta,\eta^\prime \to 0}{{\rm FR}}{\sim}& \, 4R^2\sum_{l,m} Y^m_l(\theta,\varphi)\left[Y^m_l(\theta^\prime,\varphi^\prime)\right]^* \int\limits^{+ \infty}_{- \infty} ds     \int\limits^{+ \infty}_{0} dq\,    \dfrac{ q^2 j_l \left( q \chi\right)j_l \left( q \chi^\prime \right) e^{is \left(\eta - \eta^\prime \right)}  }{ \left(s^2 - q^2 + \dfrac{5}{4} -m^2 R^2 + i \varepsilon \right)}
\nn \\
& \times
\dfrac{- is}{\left[\cosh( \pi q)-i\sinh( \pi s)\right]} \dfrac{1}{\left\vert \Gamma\left(\dfrac{3}{4}+\dfrac{iq}{2}+\dfrac{is}{2}\right)\Gamma\left(\dfrac{3}{4}+\dfrac{iq}{2}-\dfrac{is}{2}\right) \right\vert^2 }.
\label{early-time-propagator-final-1}
\end{align}

The early-time propagator \eqref{early-time-propagator-final-1} turns out to be bounded and well-defined owing to the upper bound \eqref{Gamma-estimate}. It gives rise to  a more-or-less standard correlation between points $x,x^\prime$ lying on the same sheet of the projected spacetime $\cM^{3,1}$. More remarkably, it also gives rise to a well-defined correlation and in fact a  propagation between two points located on opposite sheets of $\cM^{3,1}$  near the BB; for larger $|\eta|$ and $|\eta^\prime|$ this is suppressed compared to the case of two points on the same sheet,  due to 
the oscillating term  $e^{is \left(\eta - \eta^\prime \right)}$.

The observation that  a scalar particle can
cross the BB agrees at least qualitatively with the  classical analysis regarding  null and timelike geodesics of Sec. \ref{Sec:geodesics}. A more detailed analysis of this fascinating result is left for future work.

\section{Conclusions}

In this paper, we elaborated in detail the scalar modes and the propagator on a quantum version of a 3+1-dimensional FLRW spacetime with BB, in the semi-classical limit. The underlying framework of matrix models provides a clean setup to work with a spacetime which is singular as a classical manifold, but  well-defined as a quantum geometry. 

The most interesting conclusion is that the physics of scalar fluctuations is perfectly well-defined even at or near the classical singularity, and it is possible to relate the pre- and post-BB eras in a meaningful way. This implies some intriguing correlation between the two sides of the BB, which remain to be worked out in detail.
The framework of matrix models provides a clear prescription how to define and compute the propagator, and the causal structure of a Feynman propagator is obtained automatically, at least at late times. This is remarkable, since time emerges on the same footing as space from the underlying matrix model, and  there is no a priori notion of classical time.

This work should be seen in the context of emergent spacetime and gravity within the IKKT matrix model, which is closely related to string theory \cite{Ishibashi:1996xs}.
The present background under considerations then arises as a classical solution \cite{Sperling:2019xar}, and a gravitational action arises from one-loop effects \cite{Steinacker:2021yxt}, cf. Ref. \cite{Steinacker:2010rh}.  
Fluctuations are governed by a noncommutative field theory, which for scalar fields reduces to the one elaborated here. This suggests that analogous properties can be expected for all fluctuations in the model. In particular, the resulting theory is 
expected to satisfy the standard causality and unitarity requirements in quantum field theory, given the emergence of a Feynman propagator and the  maximal supersymmetry of the underlying IKKT model.
Even though many aspects of such a theory remain to be understood, the results of the present paper should provide sufficient motivation for further studies.

The present analysis is restricted to non-interacting test particles on the background geometry.
This is of course not entirely satisfactory, since
the density of matter is expected to be singular at the BB, which would lead to modifications of the background and hence of the metric. The inclusion of such effects as well as the induced Einstein-Hilbert action would  be very desirable, but is beyond the scope of this paper.

Finally, it is interesting to point out that recent numerical simulations of the bosonic IKKT model with mass term produced evidence for an emergent spacetime,
with  features reminiscent of a bouncing 1+1-dimensional cosmology \cite{Nishimura:2022alt}. One may therefore hope to eventually relate the matrix background under consideration to  numerical simulations of the  model.

\section*{Acknowledgement}

HS would like to thank Joanna Karczmarek for a related collaboration on the 2-dimensional case.
This work  is supported by the Austrian Science Fund (FWF) grant P32086.

\appendix
\numberwithin{equation}{section}

\section{Eigenfunctions of Laplacian operators}\label{Sec:Eigenfunctions-of-Laplacian-operators}

In this appendix, we derive the eigenfunctions of the operators $\Delta_{\mathscr{G}}$ (see Sec. \ref{Sec:Eigenfunctions-of-Delta-G}) and $\Delta^{(3)}$ (see Sec. \ref{Sec:Eigenfunctions-of-Delta-(3)}). For our convenience, we report their expressions also here:
\begin{align}
    \Delta_{\mathscr{G}} \phi &= \dfrac{1}{R^2}\Biggl\{ 3 \tanh (\eta) \partial_\eta  + \partial^2_\eta  +\dfrac{1}{\cosh^2 (\eta)} \Biggl[ \dfrac{2}{\tanh(\chi)}\partial_\chi  +\partial^2_\chi  
    \nonumber \\
   & +\dfrac{1}{\sinh^2(\chi)} \left(\dfrac{1}{\tan(\theta)}\partial_\theta +\partial^2_\theta  +\dfrac{1}{\sin^2(\theta) } \partial^2_\varphi \right) \Biggr]\Biggr\}\phi,
   \label{Laplacian-on-H4-bis}
\\
    \Delta^{(3)} \phi&= \Biggl[  \left(\dfrac{2}{\tanh \chi}\partial_\chi + \partial^2_\chi\right) + \dfrac{1}{\sinh^2 \chi} \left(\dfrac{1}{\tan \theta}
\partial_\theta + \partial^2_\theta + \dfrac{1}{\sin^2 \theta}\partial^2_\varphi\right) \Biggr]\phi. 
\label{Delta-3-without-eta-bis}
\end{align}

\subsection{Eigenfunctions of the Laplacian operator $\Delta_{\mathscr{G}}$}\label{Sec:Eigenfunctions-of-Delta-G}

In this section, we  solve the equation 
 \begin{align}
    \Delta_{\mathscr{G}} \phi = \lambda \phi,
    \label{eigenvalue-problem}
\end{align}
defining the eigenfunctions of the Laplacian operator \eqref{Laplacian-on-H4-bis}.  By exploiting the separation \emph{ansatz}
\begin{align}
    \phi(\eta,\chi,\theta,\varphi) &= \hat{f}(\eta)\hat{g}(\chi)Y^m_l(\theta,\varphi),
\label{separation-ansatz}
\end{align}
Eq. \eqref{eigenvalue-problem} gives 
\begin{align}
    \dfrac{1}{\hat{f}}\left(3 \tanh (\eta) \partial_\eta  + \partial^2_\eta \right)\hat{f} +\dfrac{1}{\hat{g}}\dfrac{1}{\cosh^2 \eta} \left(\dfrac{2}{\tanh\chi}\partial_\chi  +\partial^2_\chi\right) \hat{g} -\dfrac{l(l+1)}{\cosh^2 \eta \sinh^2 \chi} = \lambda R^2, 
\end{align}
where we have exploited that the  spherical harmonic functions $Y^m_l(\theta,\varphi)$ satisfy the property \eqref{spherical-harmonics-property}. By means of the method of  separation of variables, the above equation leads to the ordinary differential equations
\begin{subequations}
\begin{align}
\left( \partial^2_\eta + 3 \tanh (\eta) \partial_\eta -\lambda R^2 -\dfrac{\varrho}{\cosh^2 \eta} \right)\hat{f}(\eta)&=0,
\label{ode-eta}
\\
\left(\partial^2_\chi +\dfrac{2}{\tanh \chi} \partial_\chi -\dfrac{l(l+1)}{\sinh^2 \chi}  + \varrho\right)\hat{g}(\chi) &=0,
\label{ode-chi}
\end{align}
\end{subequations}
where $\varrho$ is a real constant.

Equation \eqref{ode-eta} can be put in the form of a general Legendre equation \cite{Abramowitz-Stegun(1964)}. Indeed, following the  line of reasoning of  Sec. \ref{Sec:Eigenfunctions-of-Box},  we introduce the variable
\begin{align}
    z=\tanh \eta \in (-1,1),
    \label{z-variable}
\end{align}
and write the derivative operator $\partial_\eta$  as
\begin{align}
    \partial_\eta &= (1-z^2)\partial_z.  
\end{align}
In this way, Eq. \eqref{ode-eta} becomes
\begin{align}
    \left[(1-z^2) \partial^2_z + z \partial_z - \varrho -\dfrac{\lambda R^2}{1-z^2} \right]\hat{f}(z)=0,
\end{align}
which, upon introducing
\begin{align}
    \hat{f}(z)=(1-z^2)^{3/4}\hat{h}(z),
    \label{f(z)-and-h(z)}
\end{align}
amounts to the general Legendre equation 
\begin{align}
    (1-z^2)\partial^2_z \hat{h} -2z\partial_z \hat{h} + \left[\left(\dfrac{3}{4}-\varrho\right) -\left(\dfrac{\dfrac{9}{4} +\lambda R^2}{1-z^2}\right)\right]\hat{h}=0.
\end{align}
This equation can be  solved via the associated Legendre functions of the first and second kind $\mathsf{P}^{\hat{\mu}}_{\hat{\nu}}(z)$ and $\mathsf{Q}^{\hat{\mu}}_{\hat{\nu}}(z)$, respectively, having   degree $\hat{\nu}$ and order $\hat{\mu}$ given by
\begin{subequations}
\begin{align}
\hat{\nu} &= \dfrac{1}{2}\left(2\sqrt{1-\varrho}-1\right),
\label{degree-nu}
\\
\hat{\mu} &= \dfrac{1}{2}\sqrt{9+4R^2\lambda}.
\label{order-mu}
\end{align}
\label{degree-nu-&-order-mu}
\end{subequations}
Therefore,   the solution of Eq.  \eqref{ode-eta}  is (cf. Eqs. \eqref{z-variable} and \eqref{f(z)-and-h(z)})
\begin{align}
    \hat{f}(\eta)= (1-\tanh^2 \eta)^{3/4}\left[ c_1 \mathsf{P}^{\hat{\mu}}_{\hat{\nu}}(\tanh \eta) +c_2  \mathsf{Q}^{\hat{\mu}}_{\hat{\nu}}(\tanh \eta)\right],
\label{f-eta-solution}    
\end{align}
$c_1$ and $c_2$ being  integration constants. 

The solution of \eqref{ode-chi} can be written in terms of the  ordinary hypergeometric function \cite{Abramowitz-Stegun(1964)} as
\begin{align}
    \hat{g}(\chi) &= \dfrac{\left(1-\tanh^2 \chi\right)^{(1/2)(1-a+\hat{b}+\hat{c})}}{\left(\tanh\chi\right)^{3/2}} \Biggl[ c_3 \left(\tanh^2 \chi\right)^{a/2} {}_{2}\mathrm{F}_{1}\left(\hat{b},\hat{c};a;\tanh^2 \chi\right) 
\nonumber \\    
    &+ c_4 \left(-1\right)^{1-a}\left(\tanh^2 \chi\right)^{1-a/2} {}_{2}\mathrm{F}_{1}\left(1-a+\hat{b},1-a+\hat{c};2-a;\tanh^2 \chi\right)\Biggr],
\label{g-chi-solution}
\end{align}
where $c_3$ and $c_4$ are    integration constants and
\begin{align}
a & \equiv \dfrac{1}{2} \left(3+2l\right),
\nonumber \\
\hat{b} & \equiv \dfrac{1}{2}  \left(1+l+\sqrt{1-\varrho}\right),
\nonumber \\
\hat{c} & \equiv \dfrac{1}{2}  \left(2+l+\sqrt{1-\varrho}\right) = \hat{b} + \dfrac{1}{2}.
\label{a-b-c-delta-H4}
\end{align}

We can exploit the same approach as in Sec. \ref{Sec:Eigenfunctions-of-Box}  and write Eq. \eqref{g-chi-solution} in a different form. Indeed, if we introduce the variable
\begin{equation}
y=\coth\chi,
\label{y-varibale-Appendix}
\end{equation}
then Eq. \eqref{ode-chi} can be written as (cf. Eq. \eqref{partial-chi-and-partial-y})
\begin{equation}
    (1-y^2)\partial^2_y \hat{g}(y)+ \left[l(l+1)+\dfrac{\varrho}{1-y^2}\right]\hat{g}(y)=0,
\end{equation}
and admits the solution
\begin{align}
    \hat{g}(\chi)= \sqrt{\coth^2 \chi -1}  \left[c_3 \mathcal{P}^{\sqrt{1-\varrho}}_l(\coth \chi) +c_4  \mathcal{Q}^{\sqrt{1-\varrho}}_l(\coth \chi) \right],
    \label{g-chi-solution-Legendre}
\end{align}
where we have exploited the substitution  \eqref{y-varibale-Appendix}.

Therefore,  the eigenfunctions \eqref{separation-ansatz} of the Laplacian operator \eqref{Laplacian-on-H4-bis} can be expressed  in terms  the solutions \eqref{f-eta-solution} and \eqref{g-chi-solution-Legendre}. Furthermore, from Eq. \eqref{order-mu} we obtain the explicit expression of the eigenvalue $\lambda$, i.e., 
\begin{align}
  \lambda= \dfrac{4 \hat{\mu}^2 -9}{4R^2}.  
\end{align}

It is worth underling  the similarities between the eigenfunctions   \eqref{f-eta-solution},  \eqref{g-chi-solution-Legendre} of $\Delta_{\mathscr{G}}$ and \eqref{f-eta-solution-BoxG},   \eqref{g-chi-solution-Legendre-BoxG} of the d'Alembertian operator $\Box$. The only differences between the two sets of solutions involve  the degrees and the orders of the underlying  associated Legendre functions (cf. Eqs. \eqref{degree-nu-&-order-mu} and \eqref{degree-nu-&-order-mu-BoxG}). These features stem from  the relation \eqref{relation-Box-G-Delta-G}.

\subsection{Eigenfunctions of the Laplacian operator $\Delta^{(3)}$}\label{Sec:Eigenfunctions-of-Delta-(3)}

In this section, we derive the eigenfunctions of the Laplacian operator  \eqref{Delta-3-without-eta-bis}. Like in the previous section, we solve the equation
\begin{equation}
  \Delta^{(3)} \phi (\chi,\theta,\varphi)=\lambda  \phi (\chi,\theta,\varphi),
  \label{eigenvalue-problem-for-delta-3}
\end{equation}
by means of the separation \emph{ansatz} 
\begin{align}
  \phi (\chi,\theta,\varphi) = \tilde{g}(\chi) Y^m_l(\theta,\varphi),  
\end{align}
$Y^m_l(\theta,\varphi)$ being the spherical harmonic function. In this way, after a straightforward computation, from Eq. \eqref{eigenvalue-problem-for-delta-3} we obtain
\begin{align}
    \left[ \partial^2_\chi+\dfrac{2}{\tanh \chi}\partial_\chi - \dfrac{l(l+1)}{\sinh^2 \chi}-\lambda \right]\tilde{g}(\chi)=0,
    \label{ode-for-delta-3}
\end{align}
where we have exploited Eq. \eqref{spherical-harmonics-property}. The solution of the differential equation \eqref{ode-for-delta-3} reads as
\begin{align}
    \tilde{g}(\chi) &= \dfrac{\left(1-\tanh^2 \chi\right)^{(1/2)(1-a+\tilde{b}+\tilde{c})}}{\left(\tanh\chi\right)^{3/2}} \Biggl[ c_1 \left(\tanh^2 \chi\right)^{a/2} {}_{2}\mathrm{F}_{1}\left(\tilde{b},\tilde{c};a;\tanh^2 \chi\right) 
\nonumber \\    
    &+ c_2 \left(-1\right)^{1-a}\left(\tanh^2 \chi\right)^{1-a/2} {}_{2}\mathrm{F}_{1}\left(1-a+\tilde{b},1-a+\tilde{c};2-a;\tanh^2 \chi\right)\Biggr],
\label{g-chi-solution-of-delta-3}
\end{align}
where  $c_1$, $c_2$ denote the integration constants  and
\begin{align}
a & \equiv \dfrac{1}{2} \left(3+2l\right),
\nonumber \\
\tilde{b} & \equiv \dfrac{1}{2}  \left(1+l+\sqrt{1+\lambda}\right),
\nonumber \\
\tilde{c} & \equiv \dfrac{1}{2}  \left(2+l+\sqrt{1+\lambda}\right) = \tilde{b} + \dfrac{1}{2}.
\label{a-b-c-delta-3}
\end{align}

In the same way as in Secs. \ref{Sec:Eigenfunctions-of-Box} and  \ref{Sec:Eigenfunctions-of-Delta-G}, we can write the solution of \eqref{ode-for-delta-3} in terms of the associated Legendre functions. Indeed, by introducing 
\begin{equation}
    y=\coth \chi,
\end{equation}
and bearing in mind Eq. \eqref{partial-chi-and-partial-y}, we end up with
\begin{align}
    \tilde{g}(\chi)=\sqrt{\coth^2 \chi -1}  \left[c_1 \mathcal{P}^{\sqrt{1+\lambda}}_l(\coth \chi) +c_2  \mathcal{Q}^{\sqrt{1+\lambda}}_l(\coth \chi) \right].
    \label{g-chi-solution-of-delta-3-Legendre}
\end{align}

Note that, owing to Eq. \eqref{relation-Delta-G-and-Delta-3},  Eqs. \eqref{g-chi-solution-Legendre} and \eqref{g-chi-solution-of-delta-3-Legendre} have the same structure (the only difference consists in the orders of the Legendre functions).

\section{Orthogonality relations of the associated Legendre functions}
\label{Sec:Appendix-orthogonality}

In this appendix, we will provide the details of the calculations leading to the results \eqref{integral-result-1} and \eqref{integral-result-2}.  

We will adopt the following conventions:
\begin{itemize}
\item $\mathsf{P}^{\mu}_\nu(x), \mathsf{Q}^{\mu}_\nu(x)$ denote the associated Legendre functions of  the  first and second kind, respectively, with $x$ lying in the interval $(-1,1)$;
\item $\mathcal{P}^{\mu}_\nu(x), \mathcal{Q}^{\mu}_\nu(x)$ denote the associated Legendre functions of  the first and second kind, respectively, with $x$ lying in the interval $(1,+\infty)$.
\end{itemize}

In Ref. \cite{Bielski2013}, it has been shown that 
\begin{align}
\int^1_{-1} \dfrac{dx}{1-x^2}\mathsf{P}^{is^\prime}_\nu(x) \mathsf{P}^{is}_\nu(x) &= -\dfrac{2 \Gamma(is)\Gamma(-is)}{\Gamma(1+\nu-is)\Gamma(-\nu-is)}\sin(\pi \nu)\delta(s-s^\prime)
\nonumber \\
&+\Biggl[\dfrac{\pi}{\Gamma(1-is)\Gamma(1+is)}+\dfrac{\sin^2(\pi \nu)\Gamma(is)\Gamma(-is)}{\pi}
\nonumber \\
&+ \dfrac{\pi \Gamma(is)\Gamma(-is) }{\Gamma(1+\nu-is)\Gamma(-\nu-is) \Gamma(1+\nu+is)\Gamma(-\nu+is)}\Biggr]\delta(s+s^\prime),
\end{align}
which, upon exploiting the  relations  \cite{Abramowitz-Stegun(1964)}
\begin{subequations}
\begin{align}
\Gamma (is)\Gamma (-is)&= \dfrac{\pi}{s \sinh(\pi s)},
 \label{Gamma-identity-1}
 \\
\Gamma (1-is)\Gamma (1+is)&= \dfrac{\pi s}{\sinh(\pi s)},
 \\
\Gamma (1+\nu-is)\Gamma (-\nu+is)&= \dfrac{\pi }{\sin\left[\pi(-\nu+i s)\right]},
 \\
\Gamma (1+\nu+is)\Gamma (-\nu-is)&= \dfrac{\pi }{\sin\left[\pi(-\nu-i s)\right]},
\end{align}
\end{subequations}
can be simplified, yielding
\begin{align}
\int^1_{-1} \dfrac{dx}{1-x^2}\mathsf{P}^{is^\prime}_\nu(x) \mathsf{P}^{is}_\nu(x) 
&= -\dfrac{2 \pi \sin(\pi \nu)}{s \sinh(\pi s)}\dfrac{1}{\Gamma(1+\nu-is)\Gamma(-\nu-is)}\delta(s-s^\prime)
\nonumber \\
&+2\left[\dfrac{\sinh(\pi s)}{s}+\dfrac{\sin^2(\pi \nu)}{s \sinh(\pi s)}\right]\delta(s+s^\prime).
\label{integral-Legendre-P}
\end{align}

Following the same strategy as in Ref. \cite{Bielski2013},  we will now show  that 
\begin{align}
    \int^{+\infty}_{1} \dfrac{dx}{1-x^2}\mathcal{Q}^{iq^\prime}_\nu(x) \mathcal{Q}^{iq}_\nu(x)= -\dfrac{\left(\pi/2\right)^2}{q \sinh \left(\pi q\right)} \delta(q+q^\prime),
\label{integral-Legendre-Q-1}    
\end{align}
provided  $\nu >-3$.

The explicit form of $\mathcal{Q}^{\mu}_\nu(x)$ reads either as \cite{Olver1997}
\begin{align} \label{Legendre-Q-expression-1}
    \mathcal{Q}^{\mu}_{\nu}\left(x\right)=\frac{e^{i \pi \mu }\sqrt{\pi}\Gamma\left(\nu+\mu+1%
\right)\left(x^{2}-1\right)^{\mu/2}}{2^{\nu+1}x^{\nu+\mu+1}}\mathbf{F}\left(%
\tfrac{1}{2}\nu+\tfrac{1}{2}\mu+1,\tfrac{1}{2}\nu+\tfrac{1}{2}\mu+\tfrac{1}{2}%
;\nu+\tfrac{3}{2};\frac{1}{x^{2}}\right),
\end{align}
or
\begin{align} \label{Legendre-Q-expression-2}
\mathcal{Q}^{\mu}_{\nu}\left(x\right)=\frac{2^{\nu} e^{i \pi \mu }\Gamma\left(\nu+1\right)\Gamma\left(\nu+\mu+1\right)%
(x+1)^{\mu/2}}{(x-1)^{(\mu/2)+\nu+1}}\mathbf{F}\left(\nu+1,\nu+\mu+1;2\nu+2;%
\frac{2}{1-x}\right),
\end{align}
where $\mathbf{F}\left(a,b;c;x\right)$ denotes the scaled (or Olver's) hypergeometric function, which is defined by
\begin{align}
   \mathbf{F}\left(a,b;c;x\right) = \dfrac{1}{\Gamma(a)\Gamma(b)} \sum_{n=0}^{+\infty}\dfrac{\Gamma(a+n)\Gamma(b+n)}{\Gamma(c+n)}\dfrac{x^n}{n!} = \dfrac{{}_{2}\mathrm{F}_{1}\left(a,b;c;x\right)}{\Gamma(c)},
\end{align}
${}_{2}\mathrm{F}_{1}\left(a,b;c;x\right)$ being the hypergeometric function. Starting from Eqs. \eqref{Legendre-Q-expression-1} and \eqref{Legendre-Q-expression-2}, one finds the following asymptotic approximations \cite{Olver1997}:
\begin{align}
\mathcal{Q}^\mu_\nu(x) &\overset{x \to + \infty}{\sim} \dfrac{e^{i \pi \mu} \sqrt{\pi}}{2^{\nu +1}} \dfrac{\Gamma(\nu + \mu + 1)}{\Gamma(\nu + 3/2)} \dfrac{1}{x^{\nu +1}} + {\rm O}\left(x^{-\nu -3}\right), 
\label{asymptpotic-Legendre-Q-infinity}
\\
\mathcal{Q}^\mu_\nu(x) & \overset{x \to 1^{+}}{\sim} \dfrac{e^{i \pi \mu}}{2} \Gamma(\mu) \left(\dfrac{x+1}{x-1}\right)^{\mu/2} \left[1+{\rm O}\left(x-1\right)\right].
\label{asymptpotic-Legendre-Q-1plus}
\end{align}

Given the above premises, we can now prove Eq. \eqref{integral-Legendre-Q-1}. First of all,  the general   Legendre differential equation \cite{Abramowitz-Stegun(1964)}
\begin{align}\label{general-form-Legendre-diff-eq}
    \dfrac{d}{dx}\left[(1-x^2)\dfrac{df(x)}{dx}\right] + \left[\nu(\nu+1) -\dfrac{\mu^2}{1-x^2}\right]f(x)=0,
\end{align}
written for the associated Legendre functions $\mathcal{Q}^{iq}_{\nu}(x)$  and $\mathcal{Q}^{iq^\prime}_{\nu}(x)$ having a  purely imaginary order  yields
\begin{align}\label{Legendre-diff-eq-q}
    \dfrac{d}{dx}\left[(1-x^2)\dfrac{d\mathcal{Q}^{iq}_{\nu}(x)}{dx}\right] + \left[\nu(\nu+1) +\dfrac{q^2}{1-x^2}\right]\mathcal{Q}^{iq}_{\nu}(x)=0,
\end{align}
and 
\begin{align}\label{Legendre-diff-eq-q-prime}
    \dfrac{d}{dx}\left[(1-x^2)\dfrac{d\mathcal{Q}^{iq^\prime}_{\nu}(x)}{dx}\right] + \left[\nu(\nu+1) +\dfrac{q^{\prime 2}}{1-x^2}\right]\mathcal{Q}^{iq^\prime}_{\nu}(x)=0,
\end{align}
respectively. If we subtract  Eq. \eqref{Legendre-diff-eq-q-prime} multiplied by $\mathcal{Q}^{iq}_{\nu}(x)$ from Eq. \eqref{Legendre-diff-eq-q} multiplied by $\mathcal{Q}^{iq^\prime}_{\nu}(x)$,  we find, after integrating by part,
\begin{align}
\int^\zeta_\xi \dfrac{dx}{1-x^2}\mathcal{Q}^{iq}_{\nu}(x)\mathcal{Q}^{iq^\prime}_{\nu}(x) = \dfrac{1}{q^{\prime 2}-q^2} \left\{ (1-x^2) \left[\mathcal{Q}^{iq^\prime}_{\nu}(x) \dfrac{d}{dx} \mathcal{Q}^{iq}_{\nu}(x)  - \mathcal{Q}^{iq}_{\nu}(x) \dfrac{d}{dx} \mathcal{Q}^{iq^\prime}_{\nu}(x) \right]\right\}^{x=\zeta}_{x=\xi},
\end{align}
where $\xi,\zeta \in (1,+\infty)$, with $\xi<\zeta$. Therefore, 
\begin{align} \label{integral-Legendre-Q-2}
\int^{+\infty}_1 \dfrac{dx}{1-x^2}\mathcal{Q}^{iq}_{\nu}(x)\mathcal{Q}^{iq^\prime}_{\nu}(x) &= \lim_{\zeta \to + \infty} \dfrac{1-\zeta^2}{q^{\prime 2}-q^2} \left[ \mathcal{Q}^{iq^\prime}_{\nu}(\zeta) \dfrac{d}{d \zeta} \mathcal{Q}^{iq}_{\nu}(\zeta)  - \mathcal{Q}^{iq}_{\nu}(\zeta) \dfrac{d}{d \zeta} \mathcal{Q}^{iq^\prime}_{\nu}(\zeta) \right]
\nonumber \\
&- \lim_{\xi \to 1^+} \dfrac{1-\xi^2}{q^{\prime 2}-q^2} \left[ \mathcal{Q}^{iq^\prime}_{\nu}(\xi) \dfrac{d}{d \xi} \mathcal{Q}^{iq}_{\nu}(\xi)  - \mathcal{Q}^{iq}_{\nu}(\xi) \dfrac{d}{d \xi} \mathcal{Q}^{iq^\prime}_{\nu}(\xi) \right].
\end{align}
The  limit involving the $\zeta$ variable can be easily evaluated.  Indeed,  it follows from Eq. \eqref{asymptpotic-Legendre-Q-infinity} that
\begin{align}
\lim_{\zeta \to + \infty} & \dfrac{1-\zeta^2}{q^{\prime 2}-q^2} \left[ \mathcal{Q}^{iq^\prime}_{\nu}(\zeta) \dfrac{d}{d \zeta} \mathcal{Q}^{iq}_{\nu}(\zeta)  - \mathcal{Q}^{iq}_{\nu}(\zeta) \dfrac{d}{d \zeta} \mathcal{Q}^{iq^\prime}_{\nu}(\zeta) \right]
\nonumber \\
&= \lim_{\zeta \to + \infty}  \dfrac{1-\zeta^2}{q^{\prime 2}-q^2} \left[0 + {\rm O}\left(\zeta^{-\nu-5}\right)\right] =0 , \qquad (\mbox{if }\nu >-3).    
\label{limit-integral-1}
\end{align}
For the second limit occurring in  Eq. \eqref{integral-Legendre-Q-2}, we have, after having employed Eq. \eqref{asymptpotic-Legendre-Q-1plus},  
\begin{align}
- \lim_{\xi \to 1^+} & \dfrac{1-\xi^2}{q^{\prime 2}-q^2} \left[ \mathcal{Q}^{iq^\prime}_{\nu}(\xi) \dfrac{d}{d \xi} \mathcal{Q}^{iq}_{\nu}(\xi)  - \mathcal{Q}^{iq}_{\nu}(\xi) \dfrac{d}{d \xi} \mathcal{Q}^{iq^\prime}_{\nu}(\xi) \right]  
\nonumber \\
&= \dfrac{i \Gamma(i q^\prime) \Gamma(i q)}{4 e^{\pi(q+q^\prime)}} \lim_{\xi \to 1^+} \dfrac{e^{-(i/2)(q+q^\prime)\log\left[(\xi-1)/(\xi+1)\right]}}{q+q^\prime} \left[1+{\rm O}\left(\xi-1\right)\right],
\label{limit-integral-2}
\end{align}
which can be worked out as follows. If we define the variable
\begin{align}
    u= -\dfrac{1}{2} \log\left(\dfrac{\xi-1}{\xi+1}\right) \overset{\xi  \to 1^+ }{\longrightarrow} + \infty,
\end{align}
 then we find
\begin{align}
\lim_{\xi \to 1^+} & \dfrac{e^{-(i/2)(q+q^\prime)\log\left[(\xi-1)/(\xi+1)\right]}}{q+q^\prime} = \lim_{u \to +\infty}  \dfrac{e^{iu\left(q+q^\prime\right)}}{q+q^\prime}
\nonumber \\
&= \lim_{u \to +\infty} \dfrac{\cos \left[u\left(q+q^\prime\right)\right]}{q+q^\prime} + i \lim_{u \to +\infty} \dfrac{\sin \left[u\left(q+q^\prime\right)\right]}{q+q^\prime}.
\label{limit-integral-3}
\end{align} 
Now, we exploit that, in the distributional sense,
\begin{align}
\lim_{u \to +\infty} \cos(ux)=0;
\end{align}
furthermore,  
\begin{align}
\lim_{u \to +\infty}  \dfrac{\sin(ux)}{x} = \dfrac{1}{2} \lim_{u \to +\infty} \dfrac{e^{iux}-e^{-iux}}{ix}= \dfrac{1}{2} \lim_{u \to +\infty} \int_{-u}^{u} e^{ix u^\prime} du^\prime = \pi \delta (x).
\end{align}
Therefore, Eq. \eqref{limit-integral-3} gives
\begin{align}
\lim_{\xi \to 1^+} & \dfrac{e^{-(i/2)(q+q^\prime)\log\left[(\xi-1)/(\xi+1)\right]}}{q+q^\prime} =i \pi \delta(q+q^\prime),
\label{limit-integral-4}
\end{align}
and hence, bearing in mind Eqs. \eqref{integral-Legendre-Q-2}--\eqref{limit-integral-2}, we find 
\begin{align}
\int^{+\infty}_1 \dfrac{dx}{1-x^2}\mathcal{Q}^{iq}_{\nu}(x)\mathcal{Q}^{iq^\prime}_{\nu}(x) = -\dfrac{\pi}{4}\dfrac{\Gamma(iq)\Gamma(iq^\prime)}{e^{\pi(q+q^\prime)}}\delta(q+q^{\prime})=-\dfrac{\pi}{4}\Gamma(iq)\Gamma(-iq)\delta(q+q^{\prime}), 
\label{integral-Legendre-Q-2a}    
\\ 
(\mbox{with }\nu >-3), \nonumber
\end{align}
which becomes
\begin{align}
\int^{+\infty}_1 \dfrac{dx}{1-x^2}\mathcal{Q}^{iq}_{\nu}(x)\mathcal{Q}^{iq^\prime}_{\nu}(x) = -\dfrac{(\pi/2)^2}{q \sinh(\pi q)}\delta(q+q^{\prime}), \qquad (\mbox{with }\nu >-3),
\label{integral-Legendre-Q-3}      
\end{align}
for any real $q,q^{\prime}$,
once we have exploited the identity \eqref{Gamma-identity-1}. 

The results \eqref{integral-Legendre-P} and \eqref{integral-Legendre-Q-3} have been exploited in the derivation of Eq. \eqref{orthogonality-relations-final}.

We conclude this section with a last crucial identity regarding the complex conjugation of the associated Legendre functions of second kind  $\mathcal{Q}^{iq}_{l}\left(x\right)$ (with $q \in \mathbb{R}$, $l>0$). These functions can also be written via   
\begin{align}
    \boldsymbol{Q}^{iq}_{l}\left(x\right)=\frac{\mathcal{Q}^{iq}_{l}\left
(x\right)}{e^{ -\pi q}\Gamma\left(iq+l+1\right)},
\label{Olver-associated-Legendre-associated}
\end{align}
where $\boldsymbol{Q}^{iq}_{l}\left(x\right)$ are referred to as Olver's associated Legendre functions and satisfy \cite{Olver1997}
\begin{align}
    \left[\boldsymbol{Q}^{iq}_{l}\left(x\right)\right]^* = \boldsymbol{Q}^{-iq}_{l}\left(x\right).
\end{align}
Starting from the  above relations, it is possible to prove that  
\begin{align}
\left[\mathcal{Q}^{iq}_l(x)\right]^* = e^{-2 \pi q} \mathcal{Q}^{-iq}_{l}(x).
 \label{star-of-Q-iq-l}   
\end{align}

\bibliographystyle{JHEP}
\bibliography{references}

\end{document}